\documentclass[twocolumn,tighten,usenames,dvipsnames,twocolappendix]{aastex63}
\usepackage[normalem]{ulem}
\usepackage{verbatim}
\usepackage{amssymb}
\usepackage{amsmath}
\usepackage{xspace}
\usepackage{graphicx}

\usepackage{amsmath,mathtools,mathrsfs}
\usepackage{hyperref}
\usepackage{afterpage}
\usepackage{booktabs}

\usepackage[encapsulated]{CJK}
\usepackage{ucs}
\usepackage[utf8x]{inputenc}
\newcommand{\cntext}[1]{\begin{CJK}{UTF8}{gbsn}#1\end{CJK}}

\def\muas{\mu{\rm as}} 
\def\mas{{\rm mas}} 
\def\uas{$\mu$as\xspace} 
\def\uv{$(u,v)$\xspace} 
\newcommand{\themis}{{\sc Themis}\xspace}

\newcommand{\C}{\texttt{C}\xspace}

\renewcommand{\L}{{\mathcal L}}

\newcommand{\VirA}{M87*\xspace}

\defcitealias{M87_PaperI}{Paper~I}
\defcitealias{M87_PaperII}{Paper~II}
\defcitealias{M87_PaperIII}{Paper~III}
\defcitealias{M87_PaperIV}{Paper~IV}
\defcitealias{M87_PaperV}{Paper~V}
\defcitealias{M87_PaperVI}{Paper~VI}

\shorttitle{The Photon Ring in \VirA}
\shortauthors{Broderick et al.}

\begin{document}
  
\title{The Photon Ring in \VirA}

\correspondingauthor{Avery E. Broderick}
\email{abroderick@perimeterinstitute.ca}

\author[0000-0002-3351-760X]{Avery E. Broderick}
\affiliation{Perimeter Institute for Theoretical Physics, 31 Caroline Street North, Waterloo, ON, N2L 2Y5, Canada}
\affiliation{Department of Physics and Astronomy, University of Waterloo, 200 University Avenue West, Waterloo, ON, N2L 3G1, Canada}
\affiliation{Waterloo Centre for Astrophysics, University of Waterloo, Waterloo, ON, N2L 3G1, Canada}

\author[0000-0002-5278-9221]{Dominic W. Pesce}
\affiliation{Center for Astrophysics $|$ Harvard \& Smithsonian, 60 Garden Street, Cambridge, MA 02138, USA}
\affiliation{Black Hole Initiative at Harvard University, 20 Garden Street, Cambridge, MA 02138, USA}

\author[0000-0003-2492-1966]{Roman Gold}
\affiliation{CP3 origins $|$ University of Southern Denmark (SDU) Campusvej 55, Odense, Denmark}

\author[0000-0003-3826-5648]{Paul Tiede}
\affiliation{Perimeter Institute for Theoretical Physics, 31 Caroline Street North, Waterloo, ON, N2L 2Y5, Canada}
\affiliation{Department of Physics and Astronomy, University of Waterloo, 200 University Avenue West, Waterloo, ON, N2L 3G1, Canada}
\affiliation{Waterloo Centre for Astrophysics, University of Waterloo, Waterloo, ON, N2L 3G1, Canada}
\affiliation{Center for Astrophysics $|$ Harvard \& Smithsonian, 60 Garden Street, Cambridge, MA 02138, USA}
\affiliation{Black Hole Initiative at Harvard University, 20 Garden Street, Cambridge, MA 02138, USA}

\author[0000-0001-9270-8812]{Hung-Yi Pu}
\affiliation{Department of Physics, National Taiwan Normal University, No. 88, Sec.4, Tingzhou Rd., Taipei 116, Taiwan, R.O.C.}
\affiliation{Institute of Astronomy and Astrophysics, Academia Sinica, 11F of Astronomy-Mathematics Building, AS/NTU No. 1, Sec. 4, Roosevelt Rd, Taipei 10617, Taiwan, R.O.C.}
\affiliation{Physics Division, National Center for Theoretical Sciences, Taipei 10617, Taiwan}

\author[0000-0003-3457-7660]{Richard Anantua}
\affiliation{Black Hole Initiative at Harvard University, 20 Garden Street, Cambridge, MA 02138, USA}
\affiliation{Center for Astrophysics $|$ Harvard \& Smithsonian, 60 Garden Street, Cambridge, MA 02138, USA}
\affiliation{Department of Physics \& Astronomy, The University of Texas at San Antonio, One UTSA Circle, San Antonio, TX 78249, USA}

\author[0000-0001-9240-6734]{Silke Britzen}
\affiliation{Max-Planck-Institut f\"ur Radioastronomie, Auf dem H\"ugel 69, D-53121 Bonn, Germany}

\author[0000-0002-4767-9925]{Chiara Ceccobello}
\affiliation{Department of Space, Earth and Environment, Chalmers University of Technology, Onsala Space Observatory, SE-439 92 Onsala, Sweden}

\author[0000-0002-2825-3590]{Koushik Chatterjee}
\affiliation{Black Hole Initiative at Harvard University, 20 Garden Street, Cambridge, 
MA 02138, USA}
\affiliation{Center for Astrophysics | Harvard \& Smithsonian, 60 Garden Street, Cambridge, 
MA 02138, USA}

\author{Yongjun Chen (\cntext{陈永军})}
\affiliation{Shanghai Astronomical Observatory, Chinese Academy of Sciences, 80 Nandan Road, Shanghai 200030, People's Republic of China}
\affiliation{Key Laboratory of Radio Astronomy, Chinese Academy of Sciences, Nanjing 210008, People's Republic of China}

\author[0000-0003-2886-2377]{Nicholas S. Conroy}
\affiliation{Department of Astronomy, University of Illinois at Urbana-Champaign, 1002 West
Green Street, Urbana, IL 61801, USA}
\affiliation{Center for Astrophysics | Harvard \& Smithsonian, 60 Garden Street, Cambridge, 
MA 02138, USA}

\author[0000-0002-2079-3189]{Geoffrey B. Crew}
\affiliation{Massachusetts Institute of Technology Haystack Observatory, 99 Millstone Road, Westford, MA 01886, USA}

\author[0000-0002-3945-6342]{Alejandro Cruz-Osorio}
\affiliation{Institut f\"ur Theoretische Physik, Goethe-Universit\"at Frankfurt, Max-von-Laue-Stra{\ss}e 1, D-60438 Frankfurt am Main, Germany}

\author[0000-0001-6311-4345]{Yuzhu Cui (\cntext{崔玉竹})}
\affiliation{Tsung-Dao Lee Institute and School of Physics and Astronomy, Shanghai Jiao Tong University, Shanghai, 200240, China}
\affiliation{Mizusawa VLBI Observatory, National Astronomical Observatory of Japan, 2-12 Hoshigaoka, Mizusawa, Oshu, Iwate 023-0861, Japan}
\affiliation{Department of Astronomical Science, The Graduate University for Advanced Studies (SOKENDAI), 2-21-1 Osawa, Mitaka, Tokyo 181-8588, Japan}

\author[0000-0002-9031-0904]{Sheperd S. Doeleman}
\affiliation{Black Hole Initiative at Harvard University, 20 Garden Street, Cambridge, MA 02138, USA}
\affiliation{Center for Astrophysics $|$ Harvard \& Smithsonian, 60 Garden Street, Cambridge, MA 02138, USA}

\author[0000-0002-2791-5011]{Razieh Emami}
\affiliation{Center for Astrophysics $|$ Harvard \& Smithsonian, 60 Garden Street, Cambridge, MA 02138, USA}

\author[0000-0003-4914-5625]{Joseph Farah}
\affiliation{Las Cumbres Observatory, 6740 Cortona Drive, Suite 102, Goleta, CA 93117-5575, USA}
\affiliation{Department of Physics, University of California, Santa Barbara, CA 93106-9530, USA}

\author[0000-0002-1827-1656]{Christian M. Fromm}
\affiliation{Institut für Theoretische Physik und Astrophysik, Universität Würzburg, Emil-Fischer-Str. 31, D-97074 Würzburg, Germany}
\affiliation{Institut f\"ur Theoretische Physik, Goethe-Universit\"at Frankfurt, Max-von-Laue-Stra{\ss}e 1, D-60438 Frankfurt am Main, Germany}
\affiliation{Max-Planck-Institut f\"ur Radioastronomie, Auf dem H\"ugel 69, D-53121 Bonn, Germany}

\author[0000-0002-6429-3872]{Peter Galison}
\affiliation{Black Hole Initiative at Harvard University, 20 Garden Street, Cambridge, MA 02138, USA}
\affiliation{Department of History of Science, Harvard University, Cambridge, MA 02138, USA}
\affiliation{Department of Physics, Harvard University, Cambridge, MA 02138, USA}

\author[0000-0002-3586-6424]{Boris Georgiev}
\affiliation{Department of Physics and Astronomy, University of Waterloo, 200 University Avenue West, Waterloo, ON, N2L 3G1, Canada}
\affiliation{Waterloo Centre for Astrophysics, University of Waterloo, Waterloo, ON, N2L 3G1, Canada}
\affiliation{Perimeter Institute for Theoretical Physics, 31 Caroline Street North, Waterloo, ON, N2L 2Y5, Canada}

\author[0000-0001-6947-5846]{Luis C. Ho (\cntext{何子山})}
\affiliation{Department of Astronomy, School of Physics, Peking University, Beijing 100871, People's Republic of China}
\affiliation{Kavli Institute for Astronomy and Astrophysics, Peking University, Beijing 100871, People's Republic of China}

\author[0000-0001-5160-4486]{David J. James}
\affiliation{ASTRAVEO, LLC, PO Box 1668, Gloucester, MA 01931, USA}

\author[0000-0003-2847-1712]{Britton Jeter}
\affiliation{Institute of Astronomy and Astrophysics, Academia Sinica, 11F of
Astronomy-Mathematics Building, AS/NTU No. 1, Sec. 4, Roosevelt Rd, Taipei 10617, 
Taiwan, R.O.C.}

\author[0000-0002-2662-3754]{Alejandra Jimenez-Rosales}
\affiliation{Department of Astrophysics, Institute for Mathematics, Astrophysics and Particle Physics (IMAPP), Radboud University, P.O. Box 9010, 6500 GL Nijmegen, The Netherlands}

\author[0000-0002-7029-6658]{Jun Yi Koay}
\affiliation{Institute of Astronomy and Astrophysics, Academia Sinica, 11F of Astronomy-Mathematics Building, AS/NTU No. 1, Sec. 4, Roosevelt Rd, Taipei 10617, Taiwan, R.O.C.}

\author[0000-0002-4908-4925]{Carsten Kramer}
\affiliation{Institut de Radioastronomie Millim\'etrique, 300 rue de la Piscine, F-38406 Saint Martin d'H\`eres, France}

\author[0000-0002-4892-9586]{Thomas P. Krichbaum}
\affiliation{Max-Planck-Institut f\"ur Radioastronomie, Auf dem H\"ugel 69, D-53121 Bonn, Germany}

\author[0000-0002-6269-594X]{Sang-Sung Lee}
\affiliation{Korea Astronomy and Space Science Institute, Daedeok-daero 776, Yuseong-gu, Daejeon 34055, Republic of Korea}
\affiliation{University of Science and Technology, Gajeong-ro 217, Yuseong-gu, Daejeon 34113, Republic of Korea}

\author[0000-0002-3669-0715]{Michael Lindqvist}
\affiliation{Department of Space, Earth and Environment, Chalmers University of Technology, Onsala Space Observatory, SE-43992 Onsala, Sweden}

\author[0000-0003-3708-9611]{Iv\'an Martí-Vidal}
\affiliation{Departament d'Astronomia i Astrof\'{\i}sica, Universitat de Val\`encia, C. Dr. Moliner 50, E-46100 Burjassot, Val\`encia, Spain}
\affiliation{Observatori Astronòmic, Universitat de Val\`encia, C. Catedr\'atico Jos\'e Beltr\'an 2, E-46980 Paterna, Val\`encia, Spain}

\author[0000-0001-6459-0669]{Karl M. Menten}
\affiliation{Max-Planck-Institut f\"ur Radioastronomie, Auf dem H\"ugel 69, D-53121 Bonn, Germany}

\author[0000-0002-8131-6730]{Yosuke Mizuno}
\affiliation{Tsung-Dao Lee Institute and School of Physics and Astronomy, Shanghai Jiao Tong University, Shanghai, 200240, China}
\affiliation{School of Physics and Astronomy, Shanghai Jiao Tong University, 800 Dongchuan Road,  Shanghai, 200240, People's Republic of China}
\affiliation{Institut f\"ur Theoretische Physik, Goethe-Universit\"at Frankfurt, Max-von-Laue-Stra{\ss}e 1, D-60438 Frankfurt am Main, Germany}

\author[0000-0002-3882-4414]{James M. Moran}
\affiliation{Black Hole Initiative at Harvard University, 20 Garden Street, Cambridge, MA 02138, USA}
\affiliation{Center for Astrophysics $|$ Harvard \& Smithsonian, 60 Garden Street, Cambridge, MA 02138, USA}

\author[0000-0002-4661-6332]{Monika Moscibrodzka}
\affiliation{Department of Astrophysics, Institute for Mathematics, Astrophysics and Particle Physics (IMAPP), Radboud University, P.O. Box 9010, 6500 GL Nijmegen, The Netherlands}

\author{Antonios Nathanail}
\affiliation{Institut f\"ur Theoretische Physik, Goethe-Universit\"at Frankfurt, Max-von-Laue-Stra{\ss}e 1, D-60438 Frankfurt am Main, Germany}

\author[0000-0002-8247-786X]{Joey Neilsen}
\affiliation{Villanova University, Mendel Science Center Rm. 263B, 800 E Lancaster Ave, Villanova PA 19085, USA}

\author[0000-0003-1361-5699]{Chunchong Ni}
\affiliation{Department of Physics and Astronomy, University of Waterloo, 200 University Avenue West, Waterloo, ON, N2L 3G1, Canada}
\affiliation{Waterloo Centre for Astrophysics, University of Waterloo, Waterloo, ON, N2L 3G1, Canada}
\affiliation{Perimeter Institute for Theoretical Physics, 31 Caroline Street North, Waterloo, ON, N2L 2Y5, Canada}

\author[0000-0001-6558-9053]{Jongho Park}
\affiliation{Institute of Astronomy and Astrophysics, Academia Sinica, 11F of Astronomy-Mathematics Building, AS/NTU No. 1, Sec. 4, Roosevelt Rd, Taipei 10617, Taiwan, R.O.C.}
\affiliation{Korea Astronomy and Space Science Institute, Daedeok-daero 776, Yuseong-gu, Daejeon 34055, Republic of Korea}
\affiliation{EACOA Fellow}

\author{Vincent Pi\'etu}
\affiliation{Institut de Radioastronomie Millim\'etrique, 300 rue de la Piscine, F-38406 Saint Martin d'H\`eres, France}

\author[0000-0002-1330-7103]{Luciano Rezzolla}
\affiliation{Institut für Theoretische Physik, Goethe-Universität Frankfurt, Max-von-Laue-Straße 1, D-60438 Frankfurt am Main, Germany}
\affiliation{Frankfurt Institute for Advanced Studies, Ruth-Moufang-Strasse 1, D-60438 Frankfurt, Germany}
\affiliation{School of Mathematics, Trinity College, Dublin 2, Ireland}

\author[0000-0001-5287-0452]{Angelo Ricarte}
\affiliation{Black Hole Initiative at Harvard University, 20 Garden Street, Cambridge, MA 02138, USA}
\affiliation{Center for Astrophysics $|$ Harvard \& Smithsonian, 60 Garden Street, Cambridge, MA 02138, USA}

\author[0000-0002-7301-3908]{Bart Ripperda}
\affiliation{Department of Astrophysical Sciences, Peyton Hall, Princeton University, Princeton, NJ 08544, USA}
\affiliation{Center for Computational Astrophysics, Flatiron Institute, 162 Fifth Avenue, New York, NY 10010, USA}

\author[0000-0002-1334-8853]{Lijing Shao}
\affiliation{Max-Planck-Institut f\"ur Radioastronomie, Auf dem H\"ugel 69, D-53121 Bonn, Germany}
\affiliation{Kavli Institute for Astronomy and Astrophysics, Peking University, Beijing 100871, People's Republic of China}

\author[0000-0003-0236-0600]{Fumie Tazaki}
\affiliation{Mizusawa VLBI Observatory, National Astronomical Observatory of Japan, 2-12 Hoshigaoka, Mizusawa, Oshu, Iwate 023-0861, Japan}

\author[0000-0002-7114-6010]{Kenji Toma}
\affiliation{Frontier Research Institute for Interdisciplinary Sciences, Tohoku University, Sendai 980-8578, Japan}
\affiliation{Astronomical Institute, Tohoku University, Sendai 980-8578, Japan}

\author[0000-0001-8700-6058]{Pablo Torne}
\affiliation{Max-Planck-Institut f\"ur Radioastronomie, Auf dem H\"ugel 69, D-53121 Bonn, Germany}
\affiliation{Instituto de Radioastronom\'{\i}a Milim\'etrica, IRAM, Avenida Divina Pastora 7, Local 20, E-18012, Granada, Spain}

\author[0000-0002-4603-5204]{Jonathan Weintroub}
\affiliation{Black Hole Initiative at Harvard University, 20 Garden Street, Cambridge, MA 02138, USA}
\affiliation{Center for Astrophysics $|$ Harvard \& Smithsonian, 60 Garden Street, Cambridge, MA 02138, USA}

\author[0000-0002-8635-4242]{Maciek Wielgus}
\affiliation{Max-Planck-Institut f\"ur Radioastronomie, Auf dem H\"ugel 69, D-53121 Bonn, Germany}

\author[0000-0003-3564-6437]{Feng Yuan (\cntext{袁峰})}
\affiliation{Shanghai Astronomical Observatory, Chinese Academy of Sciences, 80 Nandan Road, Shanghai 200030, People's Republic of China}
\affiliation{Key Laboratory for Research in Galaxies and Cosmology, Chinese Academy of Sciences, Shanghai 200030, People's Republic of China}
\affiliation{School of Astronomy and Space Sciences, University of Chinese Academy of Sciences, No. 19A Yuquan Road, Beijing 100049, People's Republic of China}

\author[0000-0002-9774-3606]{Shan-Shan Zhao}
\affiliation{Shanghai Astronomical Observatory, Chinese Academy of Sciences, 80 Nandan Road, Shanghai 200030, People's Republic of China}

\author[0000-0002-2967-790X]{Shuo Zhang} 
\affiliation{Bard College, 30 Campus Road, Annandale-on-Hudson, NY, 12504, USA}

\begin{abstract}
We report measurements of the gravitationally lensed secondary image---the first in an infinite series of so-called ``photon rings’’---around the supermassive black hole \VirA via simultaneous modeling and imaging of the 2017 Event Horizon Telescope (EHT) observations.  The inferred ring size remains constant across the seven days of the 2017 EHT observing campaign and is consistent with theoretical expectations, providing clear evidence that such measurements probe spacetime and a striking confirmation of the models underlying the first set of EHT results.  The residual diffuse emission evolves on timescales comparable to one week.  We are able to detect with high significance a southwestern extension consistent with that expected from the base of a jet that is rapidly rotating in the clockwise direction.  This result adds further support to the identification of the jet in \VirA with a black hole spin-driven outflow, launched via the Blandford-Znajek process.  We present three revised estimates for the mass of \VirA based on identifying the modeled thin ring component with the bright ringlike features seen in simulated images, one of which is only weakly sensitive to the astrophysics of the emission region.  All three estimates agree with each other and previously reported values.  Our strongest mass constraint combines information from both the ring and the diffuse emission region, which together imply a mass-to-distance ratio of $4.20^{+0.12}_{-0.06}~\muas$ and a corresponding black hole mass of $(7.13\pm0.39)\times10^9M_\odot$, where the error on the latter is now dominated by the systematic uncertainty arising from the uncertain distance to \VirA.
\end{abstract}

\keywords{Black hole physics --- Astronomy data modeling --- Computational astronomy --- Submillimeter astronomy --- Very long baseline interferometry --- General relativity}

\section{Introduction}
\label{sec:intro}
Radio emission from the immediate vicinity of the supermassive black hole \VirA\ was used to reconstruct the first-ever image of a black hole, estimate its mass, and interpret its theoretical environment by the Event Horizon Telescope (EHT) Collaboration 
\citep[][hereafter Papers~I-VI]{M87_PaperI,M87_PaperII,M87_PaperIII,M87_PaperIV,M87_PaperV,M87_PaperVI}. 

The analyses behind these findings included multiple image-reconstruction algorithms, model fitting in the visibility domain employing various geometric shapes, and an extensive investigation of physical emission models.  The latter are based on a large library of simulations that model the source as a hot, magnetized accretion flow, that form the input for theoretical model images constructed via ray-tracing and solving the equations of radiative transfer for a thermal population of relativistic electrons emitting synchrotron radiation.  Typically, these models predict images with both a sharp ring component -- i.e., associated with the location of photon rings in the underlying spacetime -- as well as a comparatively diffuse but still compact emission structure \citepalias[see, e.g.,][]{M87_PaperV}.

Image reconstructions, model fitting to geometric shapes, and direct fitting to general relativistic magnetohydrodynamical (GRMHD) synchrotron models all yielded an inferred mass for \VirA\ of  $(6.5\pm0.7)\times10^9 M_\odot$, which agrees with stellar-dynamical mass inferences of $(6.14_{-0.62}^{+1.07})\times10^9 M_\odot$ (\citealt{Gebhardt2011}; \citetalias{M87_PaperVI}).  The error in each method was estimated based on the variable emission structure from the GRMHD-based model images.

In the meantime, novel imaging schemes have been developed that can either approximate \citep{Arras_2019,Sun_2020} or directly sample \citep{Themaging,DMC} the posterior distribution over possible image structures.  The Bayesian nature of these schemes yield meaningful posterior distributions for the images.  These posteriors permit a more rigorous characterization of the credibility of image features and a measure of image consistency.  In addition, they permit a hybrid approach that combines image reconstruction with modeling specific expected features.  This is demonstrated in \citet{Themaging}, where imaging is accomplished with a ``nonparametric'' model comprised of a rectilinear raster of control points, and additional geometric components (Gaussians, rings, etc.).  Of particular interest here is the ability to reconstruct ring features in the data beyond the diffraction limit, provided the signal-to-noise ratios ($S/N$s) of the measured complex visibilities are sufficiently high.

Such ring features are theoretically expected outcomes in black hole images due to the propagation of photons in close proximity to the black hole, and the associated strong gravitational lensing predicted by general relativity. It is useful to think of the resulting image that is measured at infinity to be composed of a {\em direct} emission component that is dominated by the typically nontrivial and uncertain astrophysical environment, plus a series of {\em ring} components that are confined to distinct narrow regions (far less influenced by the details of the overall flow structure), which arise from photons that traversed the black hole once, twice, and so on before reaching a distant observer \citep{Bardeen1973,spin}. Thus far, the \VirA\ images present the total image structure, i.e., the sum of all of those components; the diffraction-limited image reconstructions in \citetalias{M87_PaperIV} cannot distinguish the ring components from the rest of the emission directly. 

In all of the theoretical models presented in \citetalias{M87_PaperV} that were consistent with the data, the direct emission provided the majority of the flux, with a substantial minority arising from the first higher-order image component; in what follows we will refer to these as the $n=0$  and $n=1$ ``photon rings,'' respectively, despite the former often not forming a ring.  Based on inspection of the GRMHD library presented in \citetalias{M87_PaperV}, the $n=1$ ring typically contains $10\%-30\%$ of the total compact flux, depending on the model \citep[e.g.][]{Gralla_2019,Johnson_2019}.  
Here, we demonstrate that statistically preferred reconstructions are achieved when a thin ring is included in addition to the standard nonparameteric image-reconstruction component. The thin ring is then identified with the $n=1$ photon ring of the underlying spacetime.  Interpreting the separate emission components as the $n=1$ and $n=0$ photon rings provides a powerful mapping of spacetime that is far less sensitive to the astrophysical processes in the emitting region than previous analyses \citep{spin}.

In this paper we report the results of applying the hybrid image model from \citet{Themaging} to the \VirA 2017 EHT dataset.  By fitting a geometric ring model component simultaneously with a flexible image component, we are able to isolate the $n=1$ photon ring from the surrounding diffuse emission.  We demonstrate that our reconstructed emission structure is in agreement with prior EHT imaging results, and the properties of the $n=1$ ring yield novel constraints on the black hole mass with significantly reduced systematic uncertainties.  Furthermore, because the bright ring emission is effectively extracted by the geometric model component, the image component is free to capture subtler details within the remaining diffuse emission than would otherwise be possible.  We find that the removal of the bright foreground ring uncovers additional low-brightness image structures that are consistent with originating from the base of the forward jet; the observed brightness asymmetry in context with previous EHT results constraining the black hole spin orientation \citepalias{M87_PaperV} aligns with expectations for a jet driven by the Blandford-Znajek mechanism \citep{BZ77}.

The structure of the paper is as follows. In \autoref{sec:themaging} we summarize the algorithm used to reconstruct the images. We present the reconstructed, structure-agnostic images in \autoref{sec:themages} before presenting the hybrid reconstructions in \autoref{sec:rings}. In \autoref{sec:bhparams} we describe how our findings are related to the black hole parameters.  The implications of the structure and evolution of the diffuse emission within the broader context of \VirA are collected in \autoref{sec:discussion}.  We summarize and conclude in \autoref{sec:conclusions}.

\section{Imaging Algorithm Summary}
\label{sec:themaging}
We employ the forward-modeling Markov Chain Monte Carlo (MCMC) algorithm presented in \citet{Themaging} and implemented in the \themis\ analysis framework \citep{THEMIS-CODE-PAPER}.  In this scheme, the image is forward-modeled by a rectilinear grid of control points, at which the intensity is a free parameter, and between which the intensity is modeled via an approximate cubic spline.  Station gains are simultaneously modeled, and are assumed to be fixed in time over a single scan.  For the current work, this algorithm has been supplemented in four ways.

First, we have included the field of view (FOV) as two additional parameters in the underlying image models (one for each dimension in the image plane).  This change promotes the two FOVs from hyperparameters that must be manually surveyed to parameters that are continuously explored in a fashion more consistent with the general Bayesian approach employed.  It also permits efficient exploration of asymmetric FOVs, i.e., different FOVs in the two directions of the rectilinear grid.  While this flexibility makes only a small difference for the results presented here, it does enable analysis of highly asymmetric and extended systems (e.g., active galactic nuclei jets).

Second, we permit a rotation of the rectilinear grid.  This change permits the control points to optimally arrange themselves within the confines of the grid.  Typically, this freedom permits smaller grid dimensions than grids that are fixed to be oriented along the cardinal directions.  Again, this additional flexibility makes only a modest difference in the applications here.

Third, we make use of an updated set of samplers implemented within \themis.  Details on the sampler, including demonstrations, may be found in \citet{TiedeThesis}.  In summary, the improved sampler makes use of a deterministic even-odd swap tempering scheme \citep{DEO:2019} using the Hamiltonian Monte Carlo sampling kernel from the Stan package \citep{Stan:2017}.  MCMC chain convergence is assessed using standard criteria, including integrated autocorrelation time, approximate split $\hat{R}$, and visual inspection of individual traces.

Fourth, we fit to complex visibilities rather than some combination of visibility or closure amplitudes and phases.  This requires that we reconstruct the time-variable station gain phases in addition to their amplitudes.  Both are stable across a 10~minute scan.  Fitting the complex visibilities simplifies the treatment of the errors at low $S/N$: thermal errors are strictly Gaussian \citep{TMS}.  It also improves the structure of the likelihood surface to be explored, which is both smoother and has fewer modes in this case.

Fit quality is assessed using the $\chi^2$ statistic, comparing log-likelihoods, and by inspecting the distribution of residuals.

\begin{deluxetable*}{lccccccccccc}
\tablecaption{Fit Quality Assessment \label{tab:chi2}}
\tablehead{
\colhead{Day} & 
\colhead{Model\tablenotemark{a}} &
\colhead{$N_{\rm params}$} & 
\colhead{$N_{\rm data, HI}$\tablenotemark{b}} &
\colhead{$N_{\rm data, LO}$\tablenotemark{b}} &
\colhead{$N_{\rm g,HI}$\tablenotemark{b}} & 
\colhead{$N_{\rm g,LO}$\tablenotemark{b}} & 
\colhead{$\chi_{\rm HI}^2$} & 
\colhead{$\chi_{\rm LO}^2$} & 
\colhead{$\chi^2$\tablenotemark{c}} &
\colhead{$\Delta$BIC\tablenotemark{d}} &
\colhead{$\Delta$AIC\tablenotemark{e}}
}
\startdata
April 5   & I$_{5\times5}$+A          & 34 & 336 & 336 & 162 & 162 & 295.2 & 246.4 & 628.0 & -- & --\\
          & I$_{5\times5}$+A+X        & 41 & 336 & 336 & 162 & 162 & 219.6 & 196.2 & 454.7 & -127.7 & -107.4 \\
April 6   & I$_{5\times5}$+A          & 34 & 548 & 568 & 226 & 243 & 436.3 & 376.2 & 872.1 & -- & --\\
          & I$_{5\times5}$+A+X        & 41 & 548 & 568 & 226 & 243 & 356.8 & 344.0 & 756.2 &  -66.8 &  -68.9 \\
April 10  & I$_{5\times5}$+A          & 34 & 182 & 192 &  79 &  73 &  87.9 &  96.7 & 216.5 & -- & --\\
          & I$_{5\times5}$+A+X        & 41 & 182 & 192 &  79 &  73 &  80.5 &  93.2 & 194.0 &   18.9 &   35.5 \\
April 11  & I$_{5\times5}$+A          & 34 & 432 & 446 & 185 & 190 & 388.9 & 338.1 & 800.6 & -- & --\\
          & I$_{5\times5}$+A+X        & 41 & 432 & 446 & 185 & 190 & 365.6 & 313.6 & 742.6 &  -10.5 &   -8.0 \\
\enddata
\tablenotetext{a}{Model components are as follows: an $N\times M$ dimensional image raster (I$_{N\times M}$), a large-scale asymmetric Gaussian (A), and a slashed ring (X).  Detailed descriptions and of the model components and priors can be found in the main text and \autoref{app:model}.}
\tablenotetext{b}{Each complex visibility is counted as two data points ($N_{\rm data,HI/LO}$).  Similarly, each complex gain is counted as two gain parameters ($N_{\rm g,HI/LO}$).}
\tablenotetext{c}{Includes contributions from the gain priors.}
\tablenotetext{d}{Differences between the Bayesian information criterion (BIC) with and without the slashed ring.  The BIC is defined by $\chi^2 + k\ln(N)$ where $k\equiv N_{\rm params}+N_{\rm g,HI}+N_{\rm g,LO}$ is the total number of model parameters and $N\equiv N_{\rm data,HI}+N_{\rm data,LO}$.}
\tablenotetext{e}{Differences between the Akaike information criterion (AIC) with and without the slashed ring.  The AIC is defined by $\chi^2 + 2k + 2k(k+1)/(N-k-1)$, where $k$ and $N$ are defined as they are for the BIC.}
\end{deluxetable*}

\section{Image Reconstructions} 
\label{sec:themages}

Prior to applying the complete hybrid imaging model, we first perform nonhybrid image reconstructions to more easily enable comparison with previous imaging work and to provide context for subsequent interpretation of the hybrid images.

We reconstruct images for each of the four EHT observations of \VirA taken in 2017 April.  The observations were carried out using two frequency bands \citepalias{M87_PaperIII}, which we refer to as high and low band, and we image both bands simultaneously.  We assume that the image structure is shared across bands, such that a single image is produced for each day, but we permit the station gains to be completely independent among bands.  Prior to fitting, we preprocess the visibility data as described in \citet{Themaging}: we coherently time-average the complex visibilities within each scan, and we add a systematic uncertainty of 1\% in quadrature to the thermal uncertainties to account for residual calibration errors.  This additional error budget is motivated by the analysis of nonclosing errors in \citetalias{M87_PaperIII} and the same as that found in \citetalias{M87_PaperV} and \citetalias{M87_PaperVI} sufficient to produce high-quality fits.  The model we employ uses a $5 \times 5$ grid of control points to capture the image structure, and a large-scale asymmetric Gaussian (a major-axis FWHM above 0.2~\mas; see \autoref{app:model} for details) to accommodate structure seen on only the shortest baselines (again utilizing the hybrid modeling+imaging approach).  

\begin{figure}
    \centering
    \includegraphics[width=\columnwidth]{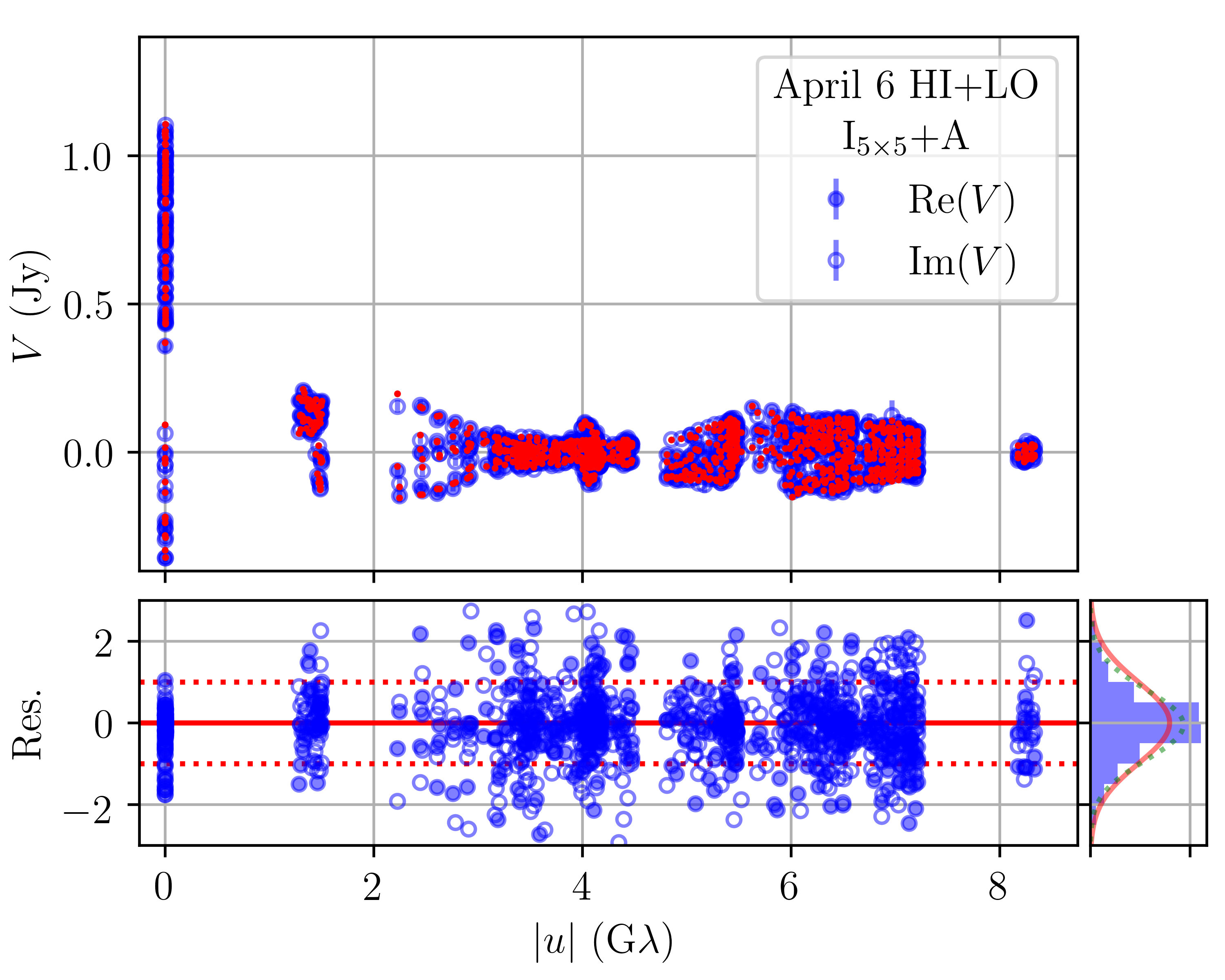}
    \includegraphics[width=\columnwidth]{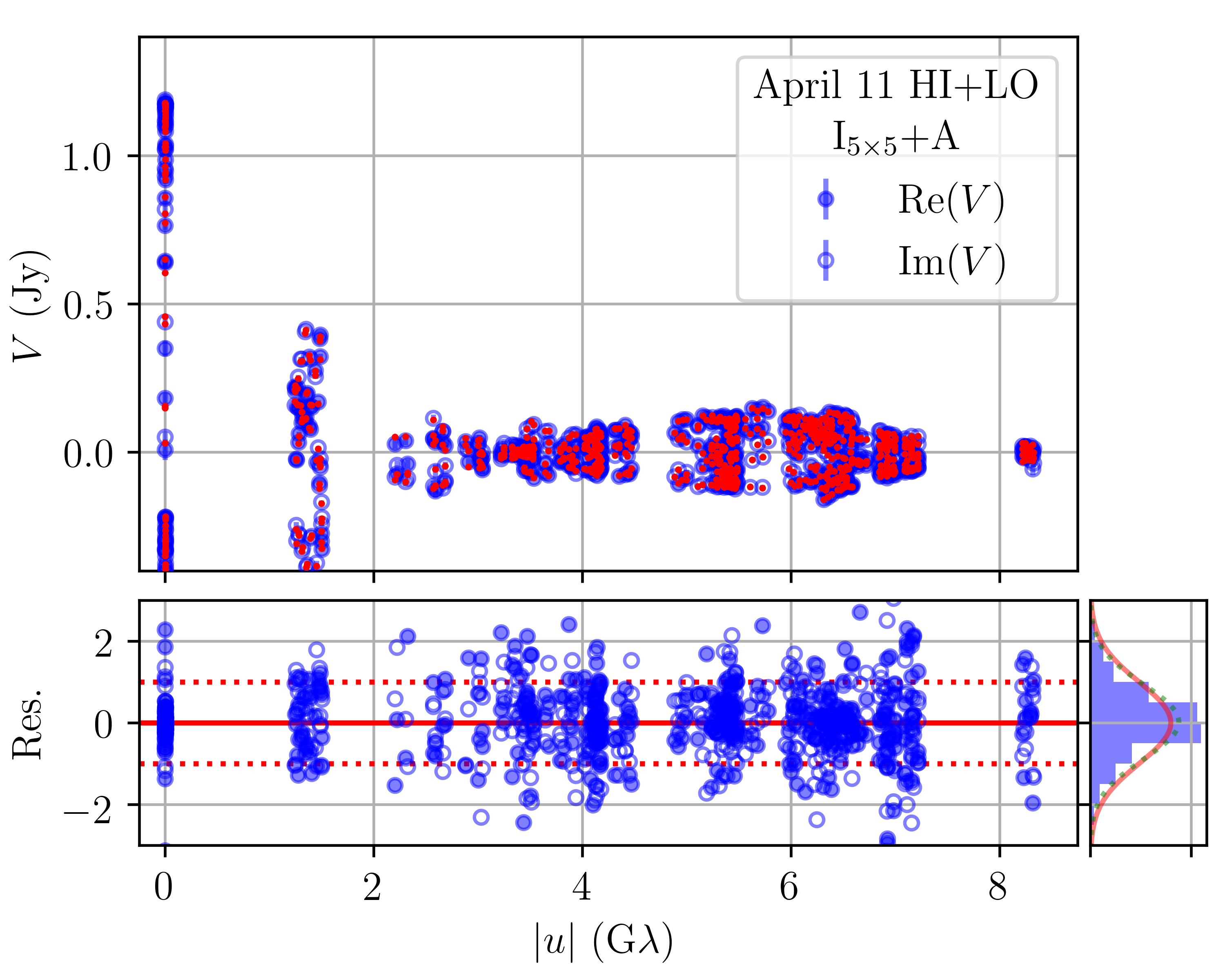}
    \caption{Direct comparisons between the model and measured values of the complex visibilities for two representative days for the combined high- and low-band data.  In each plot, the upper panel shows the maximum-likelihoood I$_{5\times5}$+A model predictions (red dots) for real (filled) and imaginary (open) components of the complex visibilities (blue points).  Normalized residuals are shown in the bottom panel with $\pm1\sigma$ indicated for reference (red dotted lines).  To the right of the residuals, the distributions of normalized residuals (blue histograms) are shown on the right in comparison to a unit-variance normal distribution (red) and the Gaussian with the same mean and variance as the residuals (green dotted).}
    \label{fig:imgres}
\end{figure}

\autoref{tab:chi2} provides an accounting of the parameters and data quantities used for each of the fits, and it lists various fit statistics for each of our reconstructions.  The quantities most comparable to the fit statistics reported in Table 5 of \citetalias{M87_PaperIV} and Table 2 of \citetalias{M87_PaperVI} would be the $\chi^2/(N_{\rm data,HI}+N_{\rm data,LO})$, which here range from 0.58 to 0.93 and are otherwise comparable to those reported previously.  Direct comparisons with the complex visibility data and the corresponding residuals are shown for representative days in \autoref{fig:imgres}.

\begin{figure*}
\begin{center}
\includegraphics[width=\textwidth]{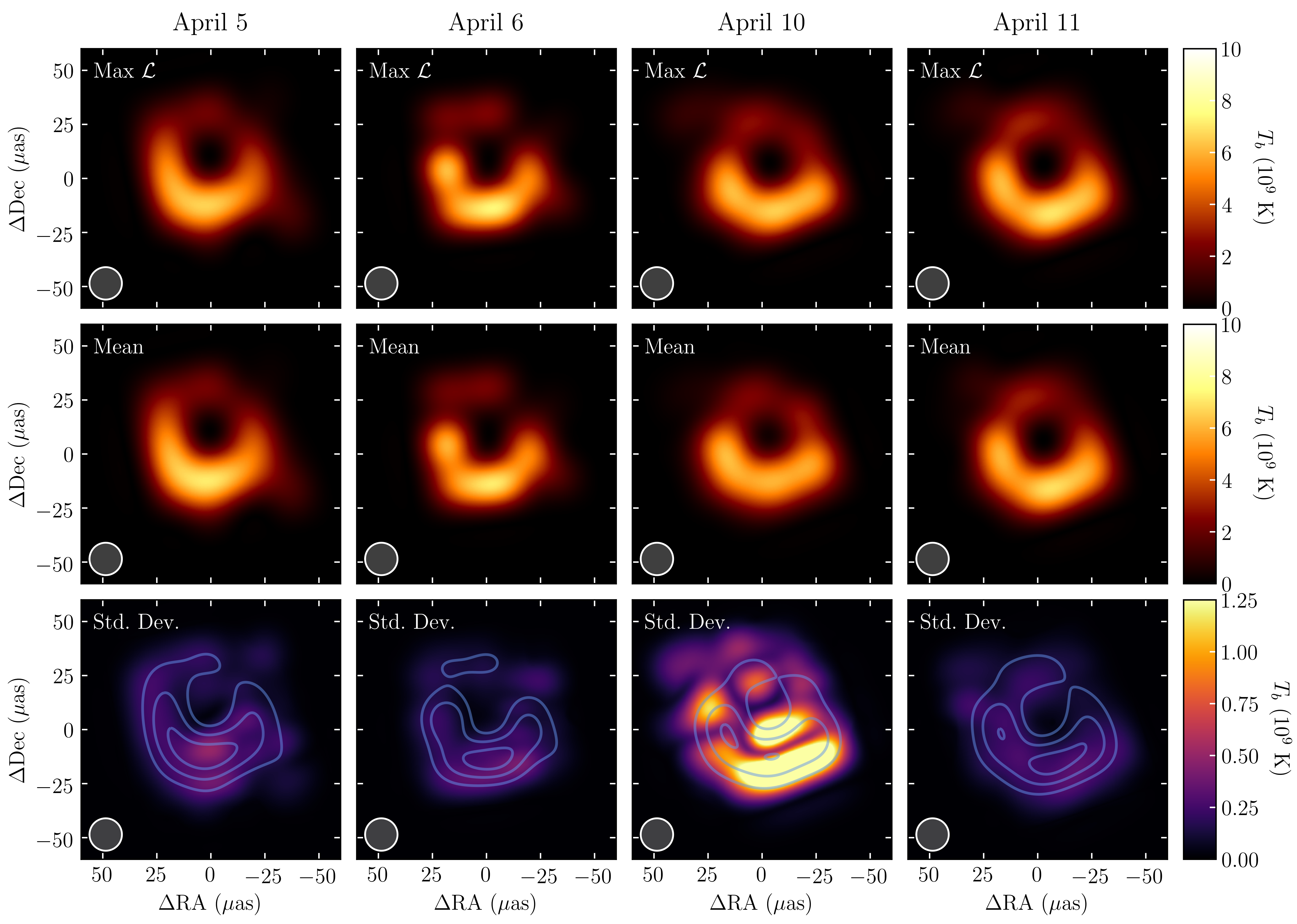}
\end{center}
\caption{Brightness temperature maps of M87 based on the raster image model.  Shown are the maximum-likelihood sample (top), average image (middle), and standard deviation with contours from the average image overlaid ranging from $2\times10^9$~K to $8\times10^9$~K in steps of $2\times10^9$~K (bottom).  Each map has been smoothed by a $15~\muas$ Gaussian beam, shown in the lower left of each panel, resulting in a combined effective resolution of approximately $20~\muas$ to make these more directly comparable with the results in \citetalias{M87_PaperIV}.}\label{fig:imgtot}
\end{figure*}

\autoref{fig:imgtot} shows the resulting image reconstructions for each of the four days.  The top row shows the maximum-likelihood samples from each chain, the middle row shows the posterior means, and the bottom row shows the posterior standard deviations.  We find that the posterior mean images show a qualitatively similar ringlike structure to image reconstructions produced using regularized maximum-likelihood (RML) methods in \citetalias{M87_PaperIV}, though we note that we have not imposed any comparable regularization (e.g., maximum entropy, total variation) on our likelihood function. The general shape, size, and total flux of the emission structure are similar across all four days, as is the pronounced north-south asymmetry in the brightness distribution. We find that the image control point raster prefers a FOV of $\sim$50\,$\muas$ in both axis directions and a modest rotation with respect to the equatorial coordinate system (see \autoref{app:model}). 

The bottom row of \autoref{fig:imgtot} illustrates a measure of the uncertainty in the image reconstructions, as previously demonstrated in \citet{Themaging}. We find that the uncertainty is not uniform across the image, nor does it seem to be proportional to the image intensity.  Rather, the uncertainty tends to be lowest within an approximately circular region running azimuthally around the ring, and it increases both radially inwards and outwards of this region.  We note that this behavior contrasts with the appearance of the image uncertainty reported in \citet{Sun_2020}; we do not explore these differences in this paper, but we expect that they arise primarily from the large differences in likelihood and prior specification between these two algorithms.

\section{Ring Reconstructions}
\label{sec:rings}

\begin{deluxetable*}{lcccc}
\tablecaption{Hybrid Image-ring Fit Parameters \label{tab:ringparams}}
\tablehead{
\colhead{Day} & 
\colhead{$I_{\rm diff}~({\rm Jy})$} &
\colhead{$\theta_{\rm diff}~(\muas)$} &
\colhead{$I_{\rm ring}~({\rm Jy})$} &
\colhead{$\theta_{\rm ring}~(\muas)$}
}
\startdata
   April 5 &   $0.252^{+0.019+0.045}_{-0.017-0.044}$ &   $17.0^{+1.1+3.4}_{-1.2-2.6}$ &   $0.301^{+0.007+0.022}_{-0.007-0.017}$ &   $21.88^{+0.13+0.36}_{-0.14-0.37}$ \\ 
   April 6 &   $0.190^{+0.014+0.037}_{-0.012-0.041}$ &   $20.2^{+1.0+2.6}_{-1.3-5.2}$ &   $0.276^{+0.010+0.049}_{-0.008-0.019}$ &   $21.44^{+0.09+0.21}_{-0.10-0.23}$ \\ 
  April 10 &   $0.176^{+0.017+0.047}_{-0.019-0.052}$ &   $20.4^{+4.2+7.0}_{-4.7-6.7}$ &   $0.302^{+0.016+0.038}_{-0.021-0.054}$ &   $21.89^{+0.27+0.51}_{-0.39-0.90}$ \\ 
  April 11 &   $0.193^{+0.013+0.035}_{-0.018-0.070}$ &   $20.8^{+0.7+2.8}_{-0.8-6.0}$ &   $0.246^{+0.020+0.058}_{-0.013-0.026}$ &   $22.51^{+0.16+0.37}_{-0.17-0.47}$ \\ 
\enddata
\tablenotetext{}{Values quoted are the median, 50-, and 90-percentile ranges.}
\end{deluxetable*}

\begin{figure}
    \centering
    \includegraphics[width=\columnwidth]{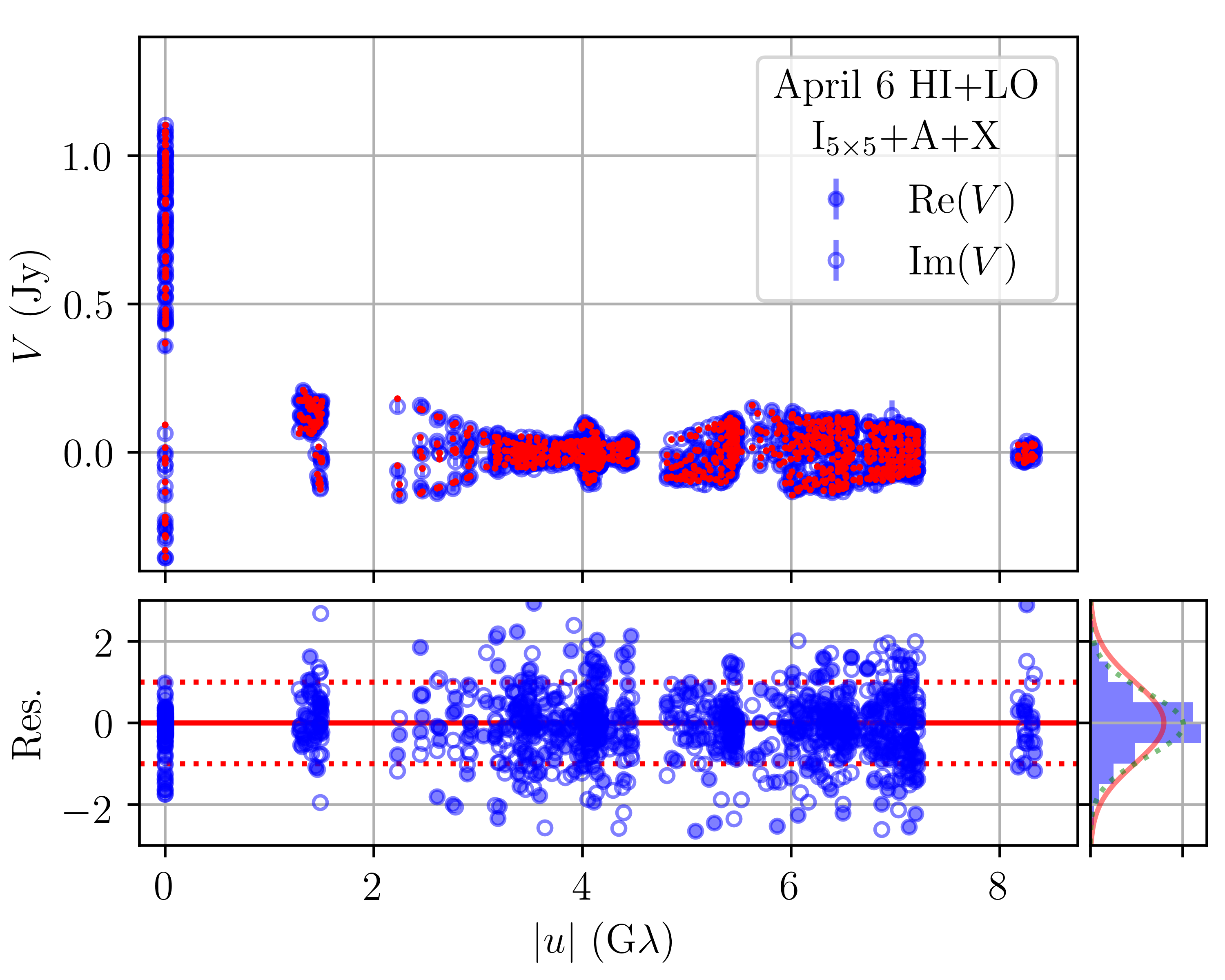}
    \includegraphics[width=\columnwidth]{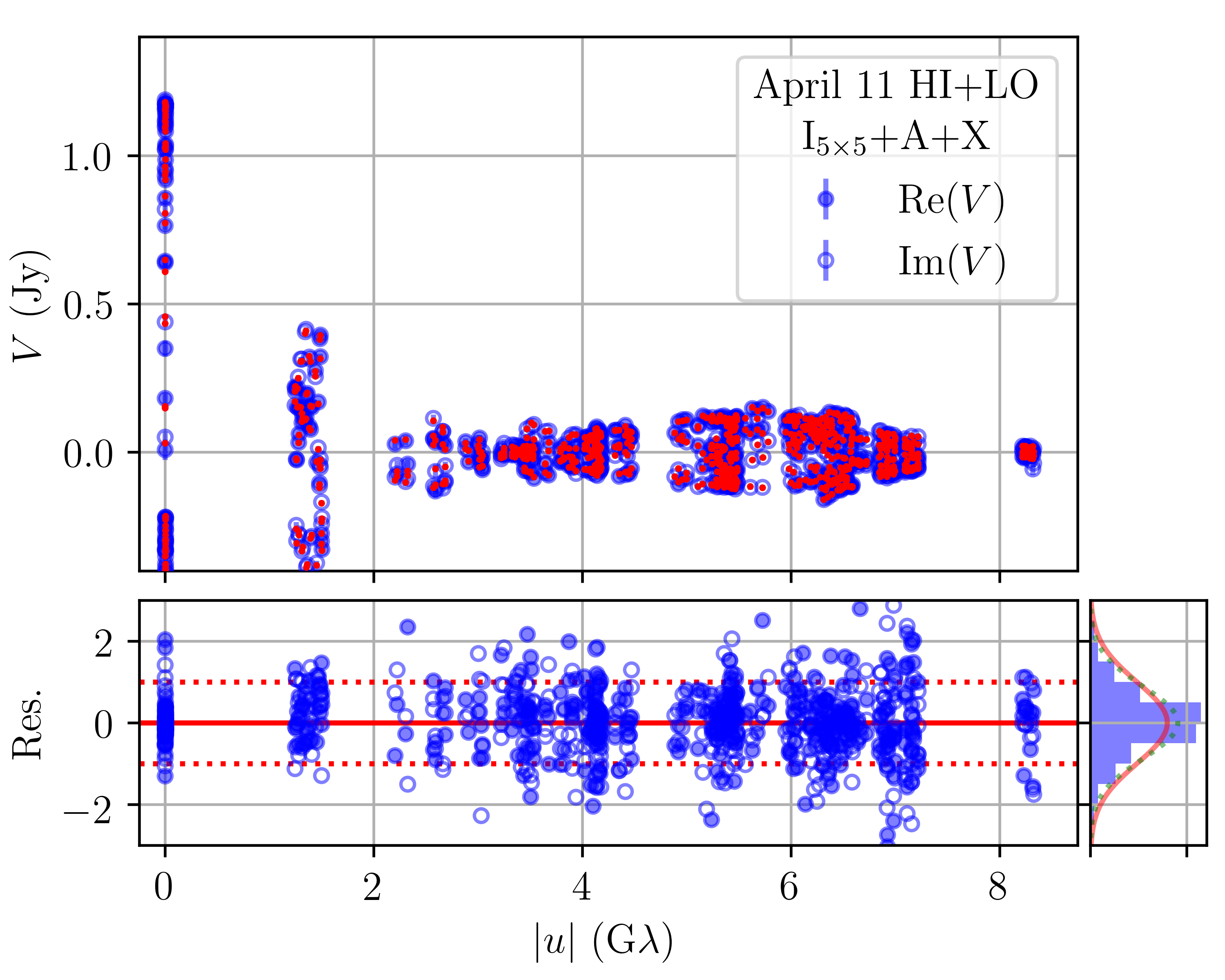}
    \caption{Similar to \autoref{fig:imgres} for the I$_{5\times5}$+A+X model.}
    \label{fig:imgreswX}
\end{figure}

\begin{figure*}
\begin{center}
\includegraphics[width=\textwidth]{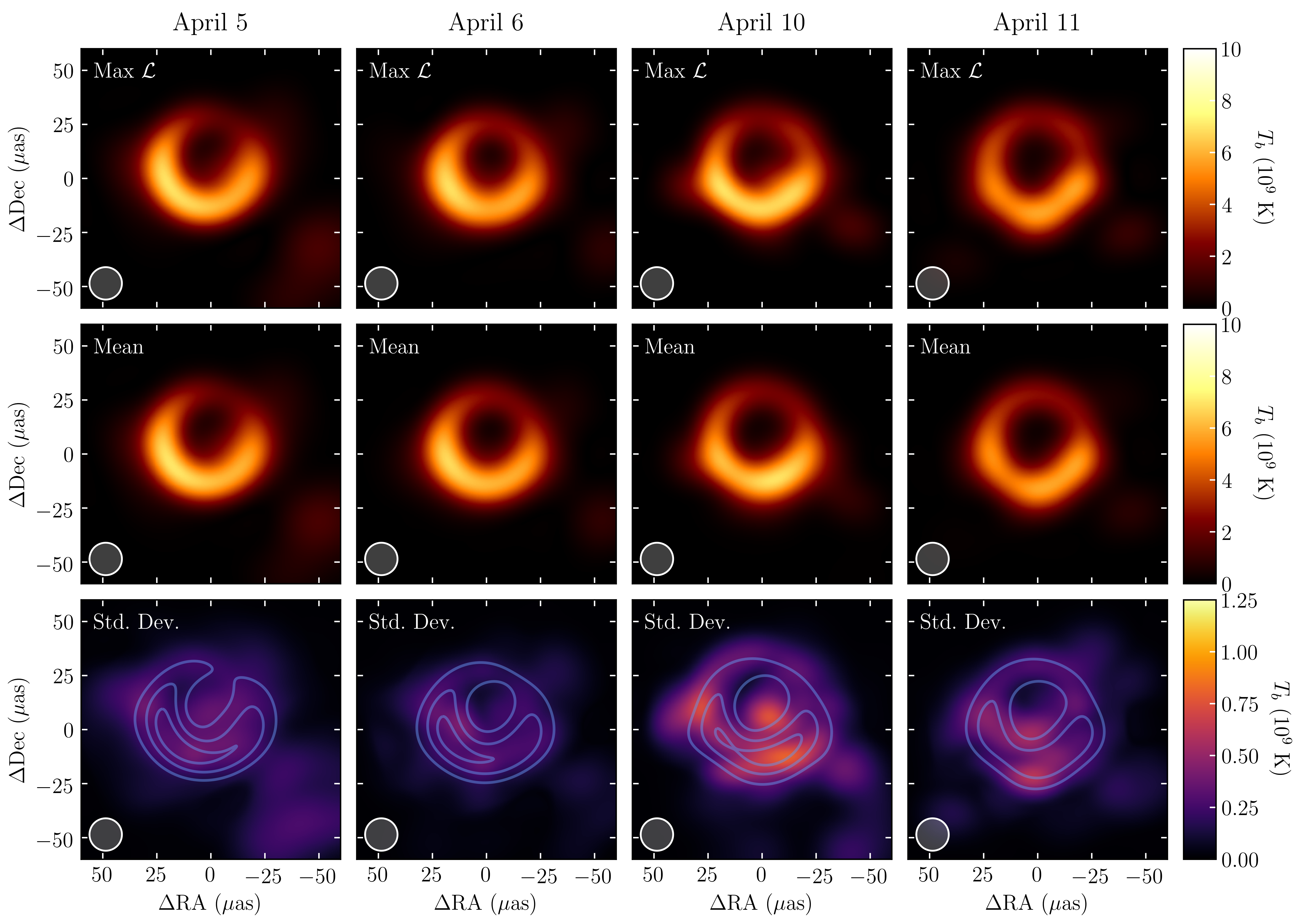}
\end{center}
\caption{Brightness temperature maps of M87 based on the hybrid ring+image model.  Shown are the maximum-likelihood sample (top), average image (middle), and standard deviation with contours from the average image overlaid ranging from $2\times10^9$~K to $8\times10^9$~K in steps of $2\times10^9$~K (bottom).  Each map has been smoothed by a $15~\muas$ Gaussian beam, shown in the lower left of each panel, resulting in a combined effective resolution of approximately $20~\muas$ to make these more directly comparable with the results in \citetalias{M87_PaperIV}.}\label{fig:imgwX}
\end{figure*}
  
The image model described in the previous section is incapable of reconstructing features on scales much smaller than the raster spacing, which is comparable to the nominal array resolution of $\sim$20\,$\muas$.  However, the lowest-order lensed image around the black hole -- i.e., the $n=1$ photon ring -- is expected to have a thickness of only $\sim$1\,$\muas$ \citep{Johnson_2019}.  While we may not expect to be able to spatially resolve the thickness of this ring with the 2017 EHT array, its $\sim$40\,$\muas$ diameter should still imprint itself on the visibility data.  By enforcing the prior expectation that a putative $n=1$ component of the observed emission structure should originate from a thin ring (i.e., one having a thickness that is much less than its diameter), \citet{Themaging} showed that the diameter of the photon ring could be reliably recovered from EHT-like synthetic datasets generated from input images produced from GRMHD simulations.

Following \citet{Themaging}, we perform hybrid imaging of the four \VirA datasets, in which we fit for a ``slashed ring'' model component alongside the image and large-scale Gaussian described in the previous section.  For the \VirA black hole, which has a spin-axis inclination of ${\lesssim}20^\circ$ with respect to the line of sight (\citealt{Walker_2018}; \citetalias{M87_PaperV}), the photon ring is expected to have a nearly circular geometry with only small ($\lesssim$2\%; \citealt{Johnson_2019}) deviations from circularity even for large spin values.  We thus model the ring as a thin (fractional thickness less than 5\%; see \autoref{app:model}) circular annulus with a linear brightness gradient (``slash''), and we permit the diameter, flux, and slash magnitude and orientation to be free parameters.  The thickness parameter is restricted by a tight prior that forces it to be $<$5\% of the diameter.  Additionally, we permit the center coordinates of the ring model component to drift with respect to the center of the image model component.  We describe the model and prior distribution specification in more detail in \autoref{app:model}.  Direct comparisons with the complex visibility data and the corresponding fit residuals for representative days are presented in \autoref{fig:imgreswX}, and indicate high-quality fits across the entirety of the range of baselines probed by the EHT.

As tabulated in \autoref{tab:chi2}, we find lower $\chi^2$ values when fitting a hybrid image to the data relative to fitting the image model described in \autoref{sec:themages}.  For comparison with Table 5 of \citetalias{M87_PaperIV} and Table 2 of \citetalias{M87_PaperVI}, the analogous $\chi^2$ quantities to those reported there range from 0.52 to 0.86, again a roughly comparable fit quality.  Such improved fit quality is expected given the increased complexity of the hybrid image model, but information criteria considerations indicate that the fit improvement outweighs the additional model complexity for all but the April 10 dataset.  The April 10 dataset is the sparsest of the four, containing a factor of ${\sim}$2--3 fewer data points than any of the other datasets, and the information criteria indicate that the increased complexity of the hybrid images is not statistically necessary in this case\footnote{Note that this does not preclude the possibility of more complex structure, as is preferred on other days.  Rather, this only indicates that more complex structures are not required to explain the data on April 10.}.  However, we note that the statistical preference for a ring component in the other three datasets---for which the coverage is more complete and the emission structure correspondingly better-constrained---implies its presence in the April 10 data set, as well.

\begin{figure*}
\begin{center}
\includegraphics[width=\textwidth]{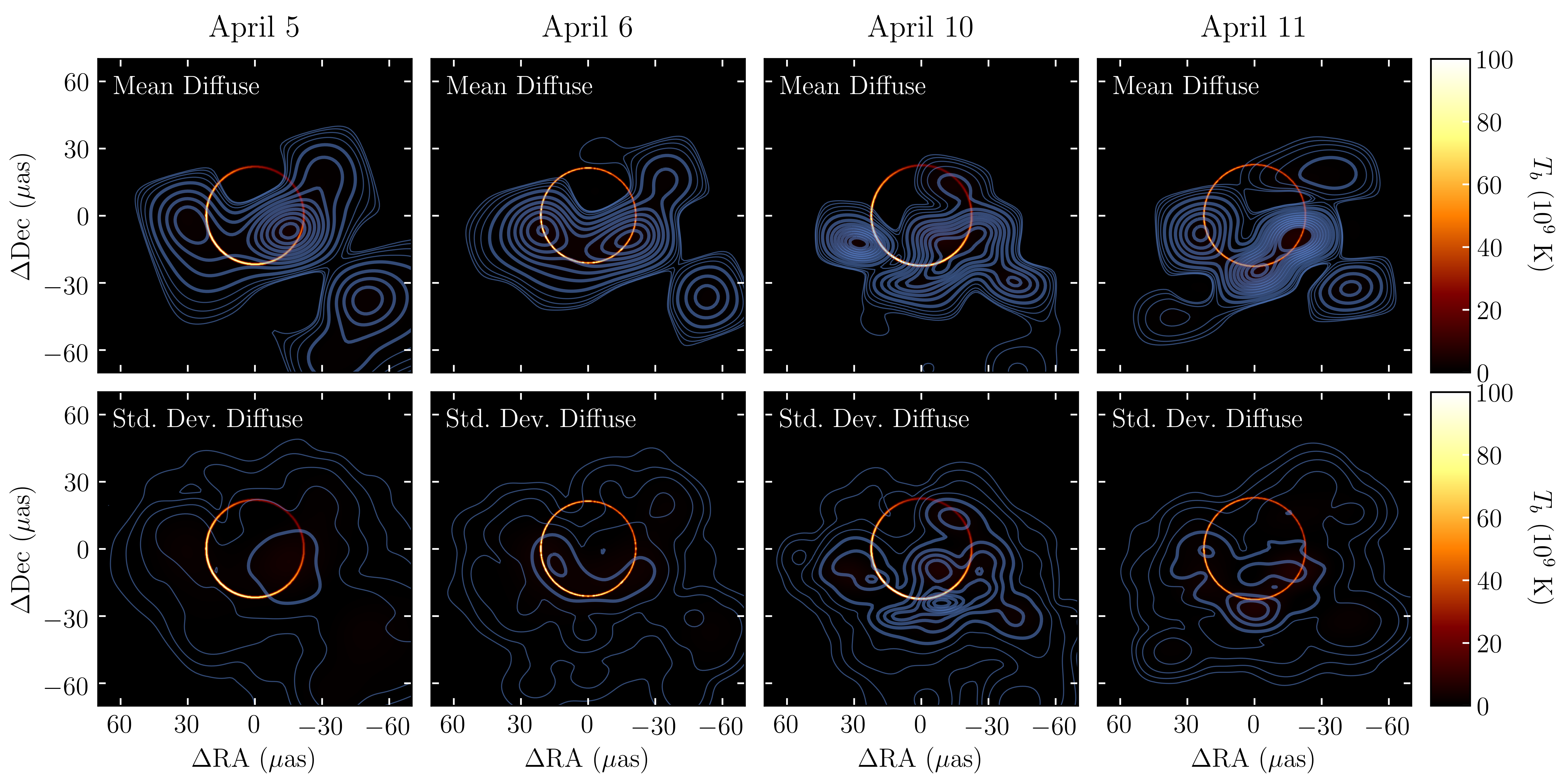}
\end{center}
\caption{Brightness temperature maps of M87 on each day with fitted slashed ring (color scale) separated from the average (top) and standard deviation (bottom) of the reconstructions of the more diffuse background image (contours) produced from $10^4$ samples drawn from the posterior. The ring color scale is linear and the ring flux map is smoothed with a circular beam with a FWHM of $0.5~\muas$.   The background emission is shown in contours --- thin contours are located at $(0.25,0.5,1,2)\times10^8$~K, thick contours are linearly spaced beginning at $4\times10^8$~K in steps of $4\times10^8$~K --- and smoothed with a circular beam with a FWHM of $15~\muas$.}\label{fig:imgsep}
\end{figure*}

Our hybrid image reconstructions for each of the four days are shown in \autoref{fig:imgwX}.  We find that after convolution with a 15\,$\muas$ Gaussian beam, the gross structural properties of the emission qualitatively match those recovered from the imaging in \autoref{sec:themages} (see \autoref{fig:imgtot}) and image reconstructions in \citetalias{M87_PaperIV}, though with an evident preference for a smoother and more uniformly circular emission structure in the hybrid images.  The ability of the thin ring model component to capture aspects of the source structure frees the remaining image model component to devote its flexibility to recovering fainter features.  This increased focus on fainter structures can also be seen in the behavior of the image control point raster, which shows a broader east-west extent for the hybrid image fits than for the image-only fits (see \autoref{app:model}).

\autoref{fig:imgsep} shows the hybrid image reconstructions with the ring and image components separated.  The slashed ring model component is shown at nearly its native resolution; it is smoothed by $0.5~\muas$ for visualization purposes only.  The diffuse emission map is displayed without any additional convolution
(in contrast to \autoref{fig:imgwX} and \autoref{fig:imgtot}).  In this figure, the slashed ring model component is visually distinct from the much-lower-brightness diffuse emission associated with the image model component.

Though the model permits both the ring and image components to drift freely with respect to one another, we find that the data prefer reconstructions in which the two components are nearly concentric.  We see that the diffuse emission is primarily concentrated along the southern portion of the ring, helping to define several ``knots'' of emission; a by-eye decomposition indicates that there are approximately two such knots on April 5 and 6, with a third knot appearing at the southernmost point of the ring on April 10 and 11.  Additionally, we find that all four days show a feature that is significantly detected to the southwest of the ring.  This southwestern component is present in some of the reconstructions from \citetalias{M87_PaperIV} and features more prominently in \citet{Arras_2020} and \citet{Carilli_2021}, but only the latter two works provided any comment; we discuss the potential origin of and implications for this feature in \autoref{sec:SWemission}.

The reconstructed flux densities in both the image and ring model components are listed in \autoref{tab:ringparams}, and we find that the ring component contains between $\sim$54\%--64\% of the total flux in the image.  This range exceeds the fraction of flux contained in the narrow ringlike features in GRMHD simulations, from which we anticipate only ${\sim}$10\%-30\% of the total image flux to be contained in the narrow ring.  The measured fluxes are, however, consistent with the results from \citet{Themaging} for hybrid image reconstructions of simulated data.  The excess ring flux appears to be a consequence of the absorption of a portion of the surrounding direct emission into the ring component.  By virtue of their sparse $(u,v)$ coverage and finite $S/N$, the EHT observations have an effective angular resolution limit of approximately ${\sim}10~\muas$ \citepalias{M87_PaperIV}, which is smaller than the nominal beam size of ${\sim}20~\muas$ but still much larger than the anticipated $n=1$ ring thickness of $\sim$1--2$~\muas$.  Image structures on scales smaller than this ${\sim}10~\muas$ threshold are effectively unresolved by the array.  Our hybrid image model priors confine the ring thickness to be $\lesssim$5\% of the ring radius (roughly $1~\muas$,; see \autoref{app:model}), but source flux contained within an annulus of thickness ${\sim}10~\muas$ around the ring radius is structurally indistinguishable from flux residing within the ring itself.  In GRMHD simulations, such an annulus contains roughly $\sim$50\%-80\% of the total source flux, consistent with the results in \autoref{tab:ringparams}.  \citet{Themaging} have demonstrated that this excess flux capture does not appear to substantially bias the recovered $n=1$ ring radius. 

\begin{figure}
\begin{center}
\includegraphics[width=\columnwidth]{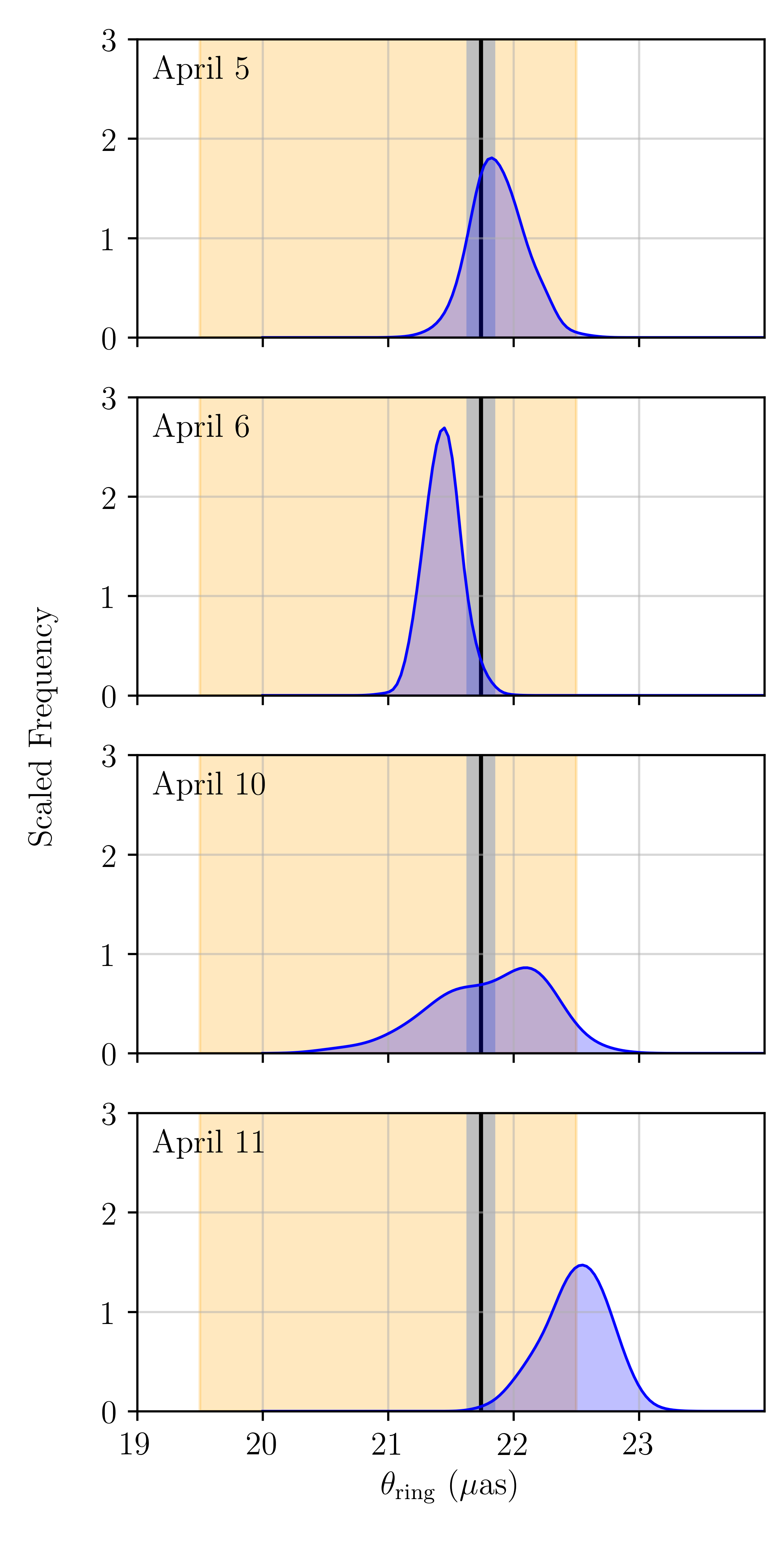}
\end{center}
\caption{Posterior distributions for the radius of the thin ring component on each day.  The variance-weighted estimates from all days and their uncertainties are indicated by the black vertical lines and gray bands, respectively.  The range of ring diameters from \citetalias{M87_PaperVI} are indicated by the orange bands.}\label{fig:RingRadii}
\end{figure}

\autoref{fig:RingRadii} shows the posterior distributions for the ring component radius parameter on all four days.  The radius measurements are consistent with being constant across the week; the apparent evolution from April 6 to April 11 is a modest $2\sigma$ tension after inclusion of the appropriate trials factor.  Combining the ring radius measurement across all days yields $\theta_{\rm ring} = 21.74\pm0.10~\muas$.  Posteriors for other ring properties (flux, position angle, thickness) can be found in \autoref{app:ancillary_posteriors}, and are similarly consistent among days. The radius of the direct emission region, $\theta_{\rm diff}$, is measured from the ring center by computing the radial location of the brightness peak in the image component.  Though we note that the direct emission does not necessarily form a closed ring on any day, this definition nevertheless presents the most consistent measure of direct emission ring size to that invoked in \citet{spin}.  We find that the direct emission radius is a factor of ${\sim}10$ more uncertain than that of the ring component; we generate posteriors for $\theta_{\rm diff}$ on each day from the hybrid image reconstructions, and the resulting values are listed in \autoref{tab:ringparams}.

\section{Physical Parameters of the Black Hole}
\label{sec:bhparams}
\begin{deluxetable}{lcc}
\tablecaption{\VirA Mass Estimates \label{tab:masses}}
\tablehead{
\colhead{Method/Origin} & 
\colhead{$\theta_M~(\muas)$} &
\colhead{$M~(10^9M_\odot)$}
}
\startdata
 \S5.1 Direct $n=1$ photon ring         & $4.15 \pm 0.74$  & $7.06 \pm 1.26$\\
 \S5.2 Corrected $n=\infty$ photon ring & $4.22 \pm 0.17$  & $7.18 \pm 0.29$ \\
 \S5.3 Joint $M$/$a$ reconstruction     & $4.20^{+0.12}_{-0.06}$  & $7.13^{+0.20}_{-0.11}$\\
 \hline
 \citetalias{M87_PaperVI} & $ 3.8\pm0.4$ & $6.5\pm0.7$\\
 \citet{Gebhardt2011} & $3.62_{-0.34}^{+0.60}$ & $6.14_{-0.62}^{+1.07}$\\
 \citet{Walsh2013}    & $2.05_{-0.16}^{+0.48}$ & $3.45_{-0.26}^{+0.85}$\\
\enddata
\tablenotetext{}{Errors indicate $1\sigma$ statistical and systematic errors, added in quadrature.  See the relevant sections for more detailed error budgets.}
\end{deluxetable}

Most directly constrained, and most comparable to other measurements of the mass, is the angular scale $\theta_M\equiv GM/c^2D$, where $D$ is the distance to \VirA. This is related to the angular radius of the bright ring.  However, it is subject to additional systematic uncertainties associated with the nature of this relationship, dominated by a systematic bias associated with the location of the emission region and the dependence on black hole spin.

Here we consider three methods for estimating $\theta_M$ and the corresponding mass, $M = M_9\times10^9 M_\odot$.  All of these identify the bright ringlike structure in the image with the $n=1$ photon ring, produced by photons that execute a half-orbit about the black hole prior to reaching the distant image plane.  This is motivated both geometrically --- higher-order photon rings are suppressed exponentially \citep{Johnson_2019} --- and from astrophysical predictions ranging from semi-analytical modeling \citep{Broderick2016} to GRMHD simulations (\citetalias{M87_PaperV}; \citealt{Porth2019,Abramowicz:2013,Gammie2003};).  They differ in the manner in which they attempt to systematically address the dependence on spin and the relationship to the asymptotic ($n=\infty$) photon ring, which defines the edge of the black hole shadow.  

In all cases, where we transform from an angular measurement of the mass, $\theta_M$ to a physical measurement, we will assume a distance of $D=16.8\pm0.8~{\rm Mpc}$ \citepalias[see Appendix I of][and references therein]{M87_PaperVI}.  These measurements and relevant comparisons are collected in \autoref{tab:masses}.

\subsection{Direct Mass Estimates}
\label{sec:mass}
As described in \citet{spin}, the sensitivity of the size of the $n=1$ photon ring to the location of an equatorial emission region is bounded.  This is in contrast to the $n=0$ photon ring, which can grow arbitrarily large with more distant equatorial emission.  As a result, with the detection of the $n=1$ photon ring, it is now possible to place a limit on the mass that is weakly dependent on the physics of the emission region, subject to the assumption that this emission is confined to near-equatorial regions, e.g., as anticipated by MAD models.\footnote{This condition may be violated by distant emission along the line of sight behind the black hole.  In such a case, however, it would be difficult to understand why the interior of the shadow does not exhibit a bright feature from a presumably foreground partner region.}

The size of the $n=1$ photon ring from equatorial emission is given by
\begin{equation}
\theta_{n=1} = \vartheta_{n=1}(a,r_{\rm em},i) \theta_M
\end{equation}
where $\vartheta_{n=1}(a,,r_{\rm em},i)$ is a dimensionless function that ranges from $4.30$ and $6.17$ for polar observers, depending on the black hole spin, $a$, and radius of the emission peak, $r_{\rm em}$.  Therefore, from the measurement of $\theta_{n=1}$ it is possible to generate an {\em astrophysics-independent} limit on the mass of \VirA:
\begin{equation}
\begin{aligned}
\theta_M
&= \vartheta_{n=1}^{-1} \theta\\
&= 4.15\pm 0.02 \pm \left. 0.74 \right|_{\vartheta_{n=1}} ~ \muas,
\end{aligned}
\end{equation}
where we have separately indicated statistical uncertainty and the systematic uncertainty in the relationship between the $n=1$ ring and the mass.  The corresponding mass estimate is
\begin{equation}
M_9 =  7.06 \pm 0.03
\pm \left.  1.26 \right|_{\vartheta_{n=1}}
\pm \left. 0.34 \right|_{D},
\end{equation}
where the systematic uncertainty associated with that on the distance is also separately stated.

\subsection{Corrected Mass Estimates}
\label{sec:grmhd_mass}
\begin{figure}
\begin{center}
\includegraphics[width=\columnwidth]{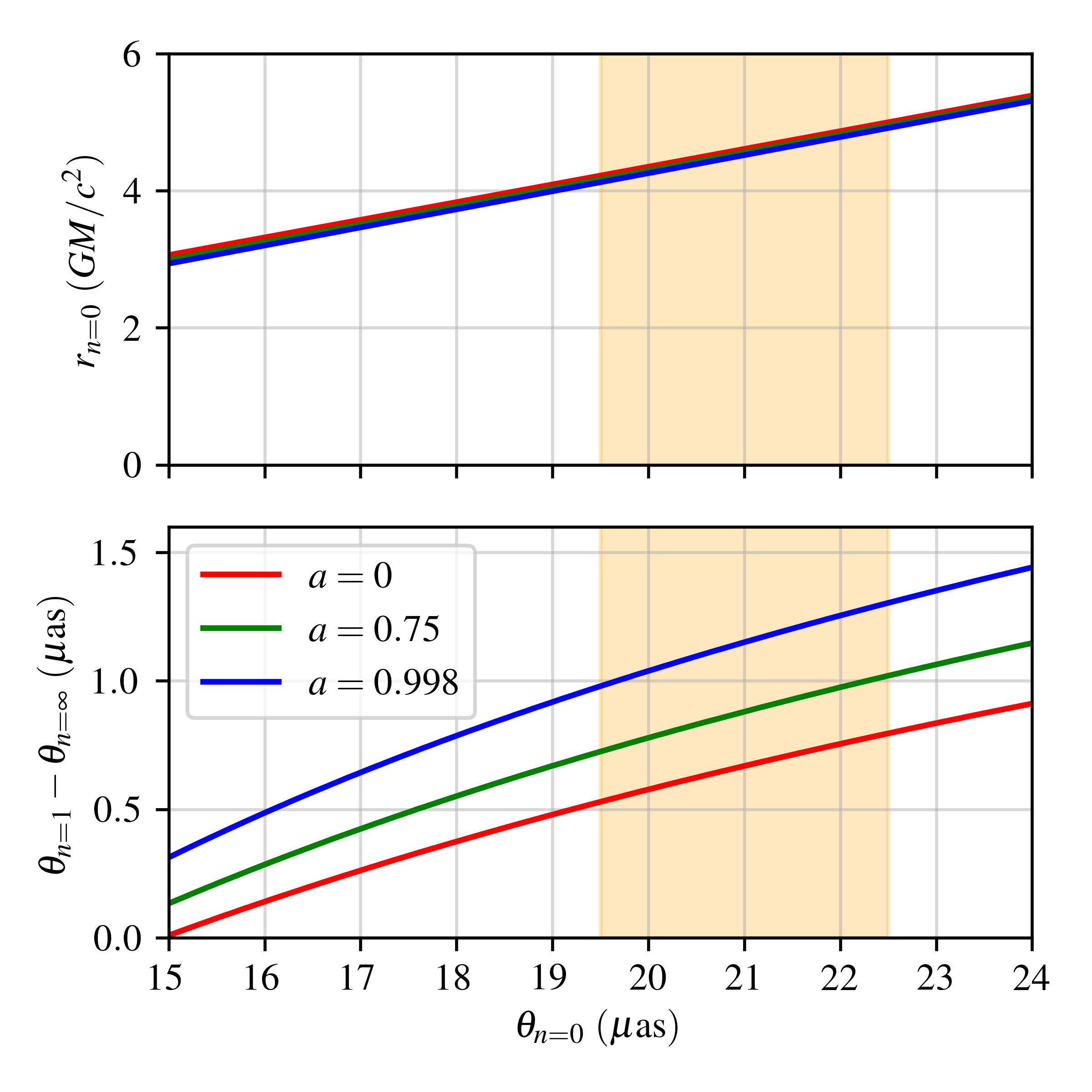}
\end{center}
\caption{Absolute shift in the radius of the $n=1$ photon ring (bottom) for a distant polar observer in comparison to the radius of the asymptotic photon ring ($n=\infty$) as functions of the apparent diameter of the direct image ($n=0$).  The orange region indicates the $1\sigma$ range of ring sizes implied by \citetalias{M87_PaperVI}, $42\pm3~\mu$as.}\label{fig:geometric_bias_summary}
\end{figure}
\begin{figure}
\begin{center}
\includegraphics[width=\columnwidth]{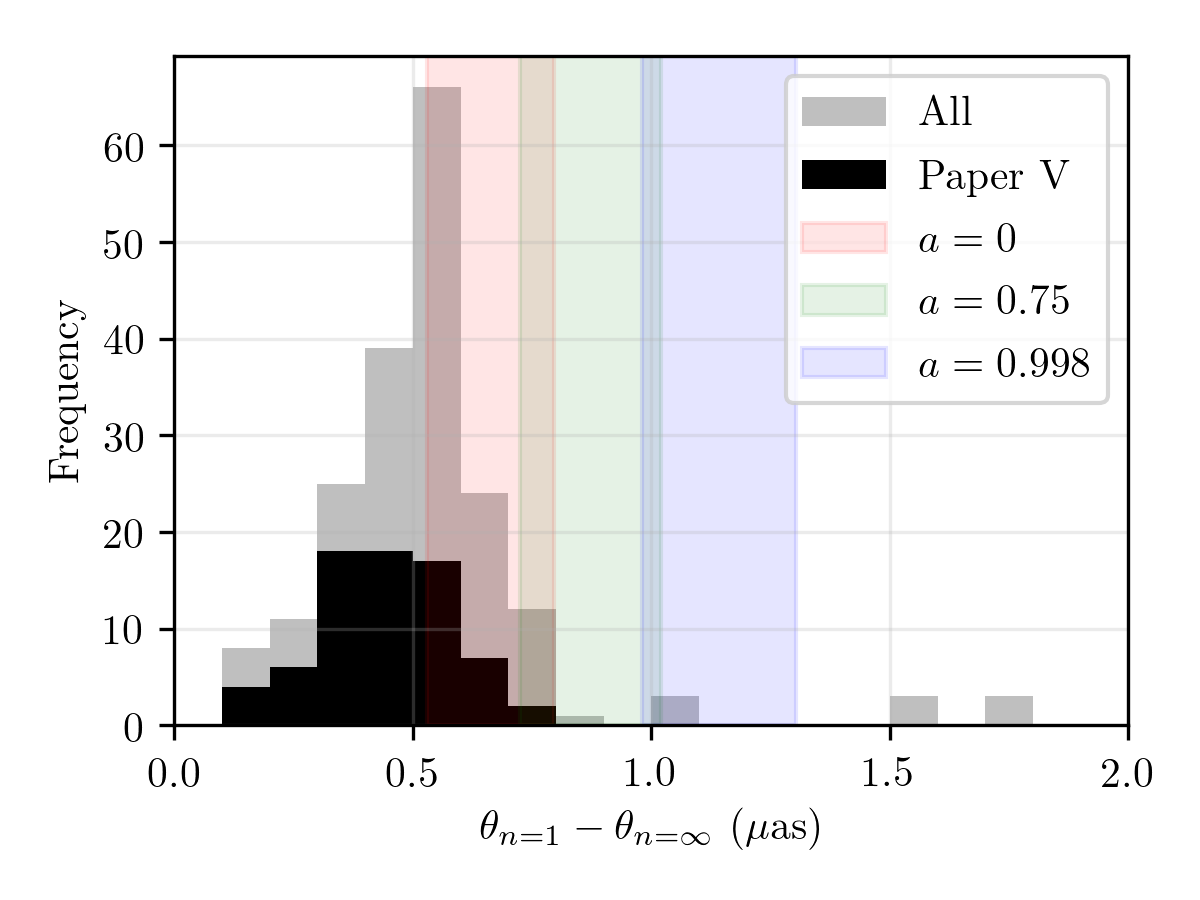}
\end{center}
\caption{Absolute shift in the radius of the $n=1$ photon ring relative to the radius of the asymptotic ($n=\infty$) photon ring from the GRMHD simulation library presented in \citepalias{M87_PaperV}, where we have identified the $n=1$ ring with the bright ringlike feature seen in the GRMHD images.  The distribution of all models \citepalias[including those excluded in][]{M87_PaperV} is shown in gray and those models that are not excluded by the 2017 EHT observations in black.  The ranges inferred from the geometric arguments, assuming emission from only the equatorial plane and using the size constraint from \citetalias{M87_PaperVI}, are shown by the vertical ranges for $a=0$ (red), $a=0.75$ (green), and $a=0.998$ (blue).}\label{fig:GRMHD_bias_summary}
\end{figure}

The $n=1$ photon ring is biased relative to the asymptotic (i.e., $n=\infty$) photon ring associated with the boundary of the black hole shadow.  The degree to which it is biased depends on the spatial distribution of the emission region \citep{Themaging,spin}.  In \autoref{app:bias} we describe two attempts to estimate the degree of this bias.  The first, described in detail in \autoref{app:geometric_bias}, is based solely on geometric arguments, assumes a polar observer, and is subject to the size constraints reported in \citetalias{M87_PaperVI}.  These are shown in \autoref{fig:geometric_bias_summary}.  These indicate that this bias is robustly limited to less than $1.3~\muas$, and typically significantly smaller.

A potentially more relevant estimate based on GRMHD simulations is described in \autoref{fig:GRMHD_bias_summary}.  We make use of the GRMHD simulations reported in \citetalias{M87_PaperV}.

These incorporate two important additional effects: the small but nonzero inclination ($i$ ranges from $12^\circ$ to $22^\circ$) and emission from above and below the equatorial plane.  Both of these tend to shrink the size of the $n=1$ photon ring, as seen in \autoref{fig:GRMHD_bias_summary}.

Models that exhibit extended emission, e.g., the $R_{\rm high}=1$ simulations, can have biases that exceed $1\,\muas$.  However, these are excluded by the \VirA size constraints \citepalias{M87_PaperV}.  When only models that are consistent with the source size and ancillary limits described in \citetalias{M87_PaperV} are considered, the size of the bias is below $0.8~\muas$; the typical shift is $\Delta\theta_{n=\infty}=0.56\pm0.32~\muas$.

Like that of the $n=1$ photon ring, the angular size of the asymptotic photon ring is given by
\begin{equation}
\theta_{n=\infty} = \vartheta_{n=\infty}(a,i) \theta_M,
\end{equation}
where $\vartheta(a,i)$ is another dimensionless function, ranging from $4.90$ to $5.20$ depending on spin for the inclinations relevant for \VirA.  This range is considerably smaller than that for $\vartheta_{n=1}$, corresponding to a significantly reduced dependence on the details of the emission region, which is otherwise encoded in $\Delta\theta$.  Thus, the uncertain spin introduces only an additional 3\% systematic uncertainty in the relationship between the asymptotic photon ring radius and the mass. 

Therefore, the mass of \VirA in angular units is given by
\begin{equation}
\begin{aligned}
\theta_M 
&= \vartheta_{n=\infty}(a,i)^{-1} \left( \theta_{n=1} - \Delta\theta_{n=\infty}\right)\\
&= 4.22 \pm 0.02
\pm \left. 0.06 \right|_{\Delta\theta}
\pm \left. 0.16 \right|_{\vartheta} ~ \muas,
\end{aligned}
\end{equation}
where again we have separately listed the random measurement error and the systematic errors associated with the emission region ($\Delta\theta$) and spin ($\vartheta$). This corresponds to a mass estimate of
\begin{equation}
M_9 =  7.18 \pm 0.04
\pm \left. 0.11 \right|_{\Delta\theta}
\pm \left. 0.27 \right|_{\vartheta}
\pm \left. 0.34 \right|_{D},
\end{equation}

\subsection{Joint Mass/Spin Estimate}
\label{sec:massspin}
Finally, following \citet{spin}, we attempt to jointly reconstruct the size of the $n=0$ and $n=1$ photon rings, as listed in \autoref{tab:ringparams}.  This leverages the additional information presented by the diffuse emission to constrain the location of the emission region.  However, because there is limited evolution in the maps during the 2017 EHT observing campaign and the simulated analyses in \citet{spin}, the constraint on spin will be weak, rendering this primarily a demonstration in principle.

\begin{figure}
\begin{center}
\includegraphics[width=\columnwidth]{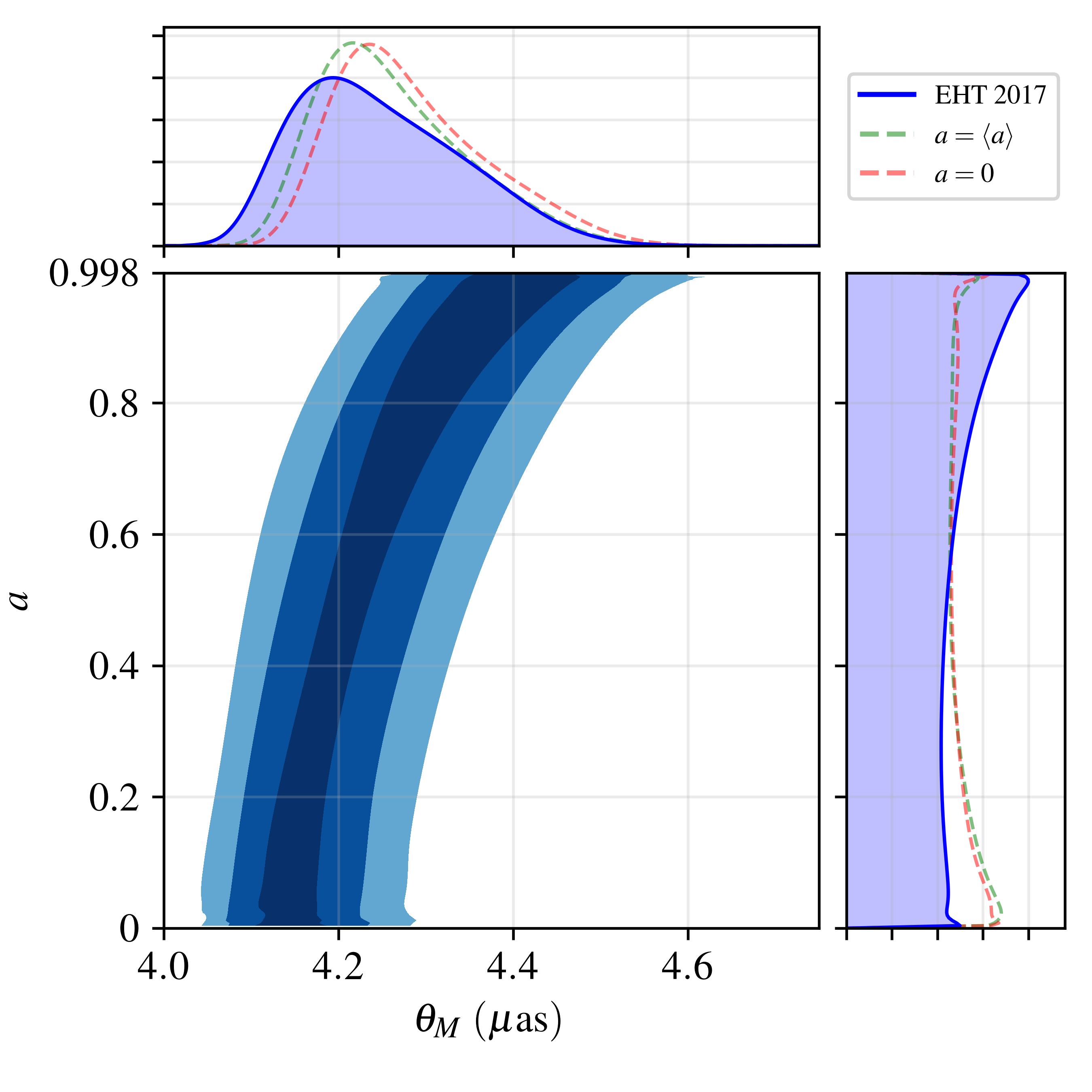}
\end{center}
\caption{Joint posterior on the spin and mass of \VirA from the measurements of $\theta_{n=0}$ and $\theta_{n=1}$ across the four observation days in the 2017 EHT campaign.  Contours indicate cumulative 50\%, 90\%, and 99\% regions.  For comparison, the one-dimensional marginalized posteriors on $M$ and $a$ are shown by the dashed lines for simulated ring sizes generated assuming a single emission radius and Gaussian errors with the sizes quoted in \autoref{tab:ringparams}.}\label{fig:Matri}
\end{figure}

Details for the model, likelihood, and method of sampling are collected in \autoref{app:massspin}.  The resulting joint posterior on $\theta_M$ and $a$ is shown in \autoref{fig:Matri}.  Marginalizing over spin, we find for the mass
\begin{equation}
\theta_M =  4.20^{+0.12+0.23}_{-0.06-0.10}~\muas,
\end{equation}
and
\begin{equation}
M_9 =  7.13^{+0.20+ 0.39}_{-0.11-0.17} \pm \left.  0.34 \right|_{D}
\end{equation}
where the indicated errors correspond to 1$\sigma$ and 2$\sigma$.  Note that because the spin is simultaneously reconstructed, these include what has previously been identified as the systematic uncertainties associated with the n=1 photon ring size bias and spin.

After marginalizing over mass, the spin estimate is
\begin{equation}
a = 1.0^{+0+0}_{-0.5-1},
\end{equation}
where again the indicated errors correspond to 1$\sigma$ and 2$\sigma$.  As anticipated, the spin is effectively unconstrained, although there appears to be a very weak preference for high spin.

To demonstrate this we repeated the analysis with synthetic size measurements constructed from the equatorial emission model with two sets of $(\theta_M,a)$.  The first was set to the average values from the joint posteriors, $(\theta_M,a)=(4.32~\muas,0.63)$.  The second was to set to $(\theta_M,a)=(4.2~\muas,0.0)$, exploring the posterior associated with a truth value of zero spin.  These are shown in \autoref{fig:Matri} by the dashed red and green lines, respectively, in the one-dimensional, marginalized posteriors for $\theta_M$ and $a$.  Both exhibit the same posterior excess near $a=1$, implying that it is not significant.

\subsection{Synthesis and Discussion}
The accuracy of previous EHT estimates of $\theta_M$ for \VirA is dominated by systematic uncertainties associated with the astrophysics of the emission region (\citetalias{M87_PaperVI}; \citealt{Gralla_2019}).  However, the detection of the bright ring reduces a number of these uncertainties, both qualitatively and quantitatively.

The mass estimate presented in \autoref{sec:mass}, in which the bright ring is identified with the $n=1$ photon ring, is independent of even pathological near-equatorial emission distributions.  As a result, the direct detection of the bright ring has effectively produced a mass estimate in which the impacts of astrophysical uncertainties are strictly bounded.  While the systematic uncertainty associated with this detection is nearly twice that reported in \citetalias{M87_PaperVI}, it is no longer dependent on the astrophysical calibration procedure used there, eliminating a key astrophysical uncertainty.

The combined image reconstructions confirm that the image morphology on each day is similar to those produced by GRMHD simulations, comprised of a bright ring and a diffuse, more variable surrounding emission structure.  This provides a strong conceptual foundation for the calibrated mass estimates using GRMHD simulations presented in \citetalias{M87_PaperVI} and revised in \autoref{sec:grmhd_mass}.  The inclusion of a prior expectation on the size of the emission region, inherent in the GRMHD simulations, results in significant improvements of systematic uncertainties.  As a result, the effective systematic uncertainty is reduced to roughly half of that in \citetalias{M87_PaperVI}.  

Finally, the diffuse emission provides a direct, astrophysics-independent estimate of the emission region location.  Joint modeling of the bright ring and the diffuse component as the $n=1$ and $n=0$ photon rings, respectively, generates an astrophysics-independent estimate of the mass that incorporates the remaining systematic uncertainties directly as statistical errors.  The half-range is approximately a quarter of that quoted in \citetalias{M87_PaperVI}.  This implies that the position of the $n=0$ photon ring provides a stronger constraint on the location of the emission region than the prior inferred from the GRMHD simulations from \citetalias{M87_PaperV}.

All of the mass estimates presented here are consistent among each other within their respective systematic errors.  They are also consistent with the combined mass estimates presented in \citetalias{M87_PaperVI}, though they lie at the high end of the mass range listed there.  This suggests that those GRMHD simulations with more compact emission regions are more consistent with the diffuse emission maps reconstructed here.

These mass estimates are also consistent with those arising at scales of $10^2$~pc from the dynamics of stars \citep{Gebhardt2011}.  It remains inconsistent with the gas dynamical mass estimate reported in \citet{Walsh2013}.  The significance of this discrepancy has now grown to more than $4\sigma$.  This inconsistency may be ameliorated by adjustments in the underlying gas disk model \citep{Jeter2019,Jeter2020b}.

\section{Origin and Evolution of the Diffuse Component}
\label{sec:discussion}
The detection of a bright ringlike feature, and its separation from the diffuse component, has a number of immediate implications for the origin of the emission and the properties of the central black hole.  As shown explicitly in \citet{Themaging}, the morphology of the diffuse emission is accurately recovered despite the absorption of excess flux into the thin ring component.  Thus, here we discuss the structure the diffuse component in the broader context of the environment and properties of \VirA.

\subsection{Evolution of M87 from 2017 April 5-11}
There are clear signatures of evolution across the EHT observation campaign.  The diffuse emission maps on neighboring days are similar.  This is consistent with expectations based on the dynamical timescales; $GM/c^3\approx9$~hr in \VirA and GRMHD simulations indicate little evolution on timescales shorter than $10 GM/c^3$ \citep{Porth2019}.  

In contrast, significant evolution in the diffuse emission occurs between the first two days (April 5, 6) and the last two days (April 10, 11).  This evolution seems to primarily manifest as the addition in the later two days of a distinct component to the southern region of the diffuse ring.  The absence of an intervening observation leaves the origin of this southern component unclear; it could be a new component appearing, or it could be a growing extension of the western component.  The latter interpretation would be consistent with the clockwise rotation of the black hole inferred from the orientation of the diffuse ring \citepalias{M87_PaperIV} and possibly to outflowing features within the jet \citep{Jeter2020}.  Note that the sense of this rotation is opposite to the predominantly counter-clockwise evolution identified in the total emission map, exhibited in \autoref{fig:imgtot} and \autoref{fig:imgwX}, as well as other analyses (\citetalias{M87_PaperIV}; \citealt{Arras_2020}; \citealt{Carilli_2021}).

Equally important, if not more so, is what does {\em not} evolve.  No significant changes in the angular size of the narrow ring component are detected.  This is consistent with the interpretation of this feature as primarily gravitational, and thus dependent only weakly on the details of the otherwise evolving emission region.

\subsection{Extended Southwestern Emission} \label{sec:SWemission}
\begin{figure*}
\begin{center}
\includegraphics[width=\textwidth]{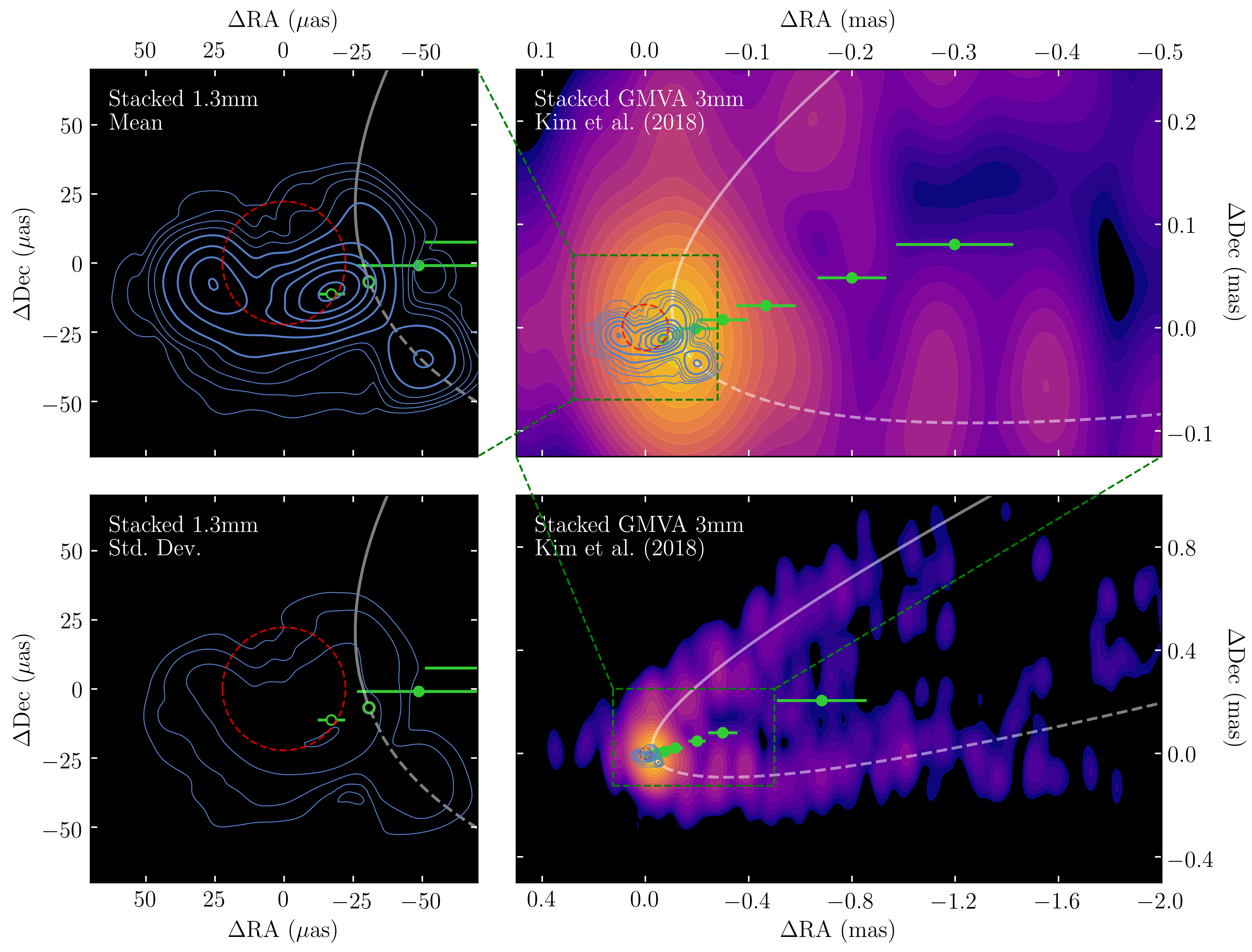}
\end{center}
\caption{Comparison of the stacked image of the diffuse emission produced by the variance-weighted mean (upper left) and its standard deviation (lower left) with the stacked GMVA map at 3~mm from \citet{M873mm} (right).  In the GMVA map, the extended jet is clearly visible, and the contours are located at $(-1,1,1.414,2,\dots)\times0.47~{\rm mJy~beam^{-1}}$; the contours of the diffuse emission map are the same as in \autoref{fig:imgsep}.  The fit to the ridgelines of the limb-brightened emission from \citet{M873mm} are shown by white solid (top) and dashed (bottom) lines in each panel, with the origin shown by the open green circle.  A ring with the mean radius of $22.22~\muas$ is indicated by the dashed red line.  Boundaries of neighboring panels are indicated in green dashed lines.  Shown by green circles are the measured core shifts at 2.3~GHz (rightmost), 5~GHz, 8.4~GHz, 15.2~GHz, 23.8~GHz and 43.2~GHz (leftmost closed)  from \citet{Hada2011} referenced to the anticipated location of the 3~mm core (right open).  The expected location of the 1.3~mm core (left open) matches the peak of the diffuse emission.}\label{fig:EHTGMVA}
\end{figure*}

In all of the diffuse emission maps shown in \autoref{fig:imgsep}, an extension to the southwest is visible.  The ability to produce a statistically meaningful image posterior enables the assignment of a significance to these features, which ranges from $4\sigma$ to $12\sigma$ on the individual days.  On a three of the four of days there is a matching northwestern extension, though this is less significant ($1\sigma$-$2\sigma$).

While such features appear similar to the dirty beams seen in some \VirA images produced by other algorithms \citepalias[see Figures~7 and 8 of][]{M87_PaperIV}, none have been seen at statistically significant levels (as characterized by the image posteriors) in the various simulated data tests performed with the Bayesian scheme employed here \citep[see, e.g., Figure~4 of][]{Themaging}.  Similar claims of a statistically significant detection have been made by other groups \citep{Arras_2020,Carilli_2021}.

It is suggestive that the orientation of these diffuse extensions align with the limb-brightened jet seen at 3~mm, shown in \autoref{fig:EHTGMVA}.  We align the center of light of the mean 1.3~mm maps, averaged over the four observation days in 2017, and the centroid of the core component of the 3~mm maps.  The ridgeline fits from \citet{M873mm} are also shown in \autoref{fig:EHTGMVA}, assuming a jet position angle of $69^\circ$ east of north and a width, $W\propto z^{-0.498}$, where $z$ is the projected core distance.  Additional core shifts along the jet of $25~\muas$ and transverse to the jet of $-10~\muas$ are applied.  
Extrapolating the core shift power law determined from longer wavelengths by \citet{Hada2011}, places the anticipated 1.3~mm core on top of the brightest diffuse component.\footnote{The positions of the 1.3~mm and 3~mm cores are strongly correlated in the fits reported in \citet{Hada2011}.  Thus, the {\em relative} positions of the 1.3~mm and 3~mm cores is much better constrained than the absolute position of either in relation to the 7~mm core.  We estimate the uncertainty in the location of the 1.3~mm core via Monte Carlo sampling of the power-law fit parameters reported in \citet{Hada2011}, assuming independent Gaussian errors in the fit parameters.}  With these shifts, the ridgelines connect from the southwestern and northwestern extensions.  We caution that the apparent structure of the jet base on horizon scales may depart substantially from the power-law behavior at large scales due to the combined effects of inclination, gravitation lensing at small radii, inhomogeneous evolution in the optical depth across the image, and relativistic motion \citep[see, e.g.,][]{Broderick2009, Moscibrodzka2016, Chael19, Davelaar19}.

\begin{figure}
\begin{center}
\includegraphics[width=\columnwidth]{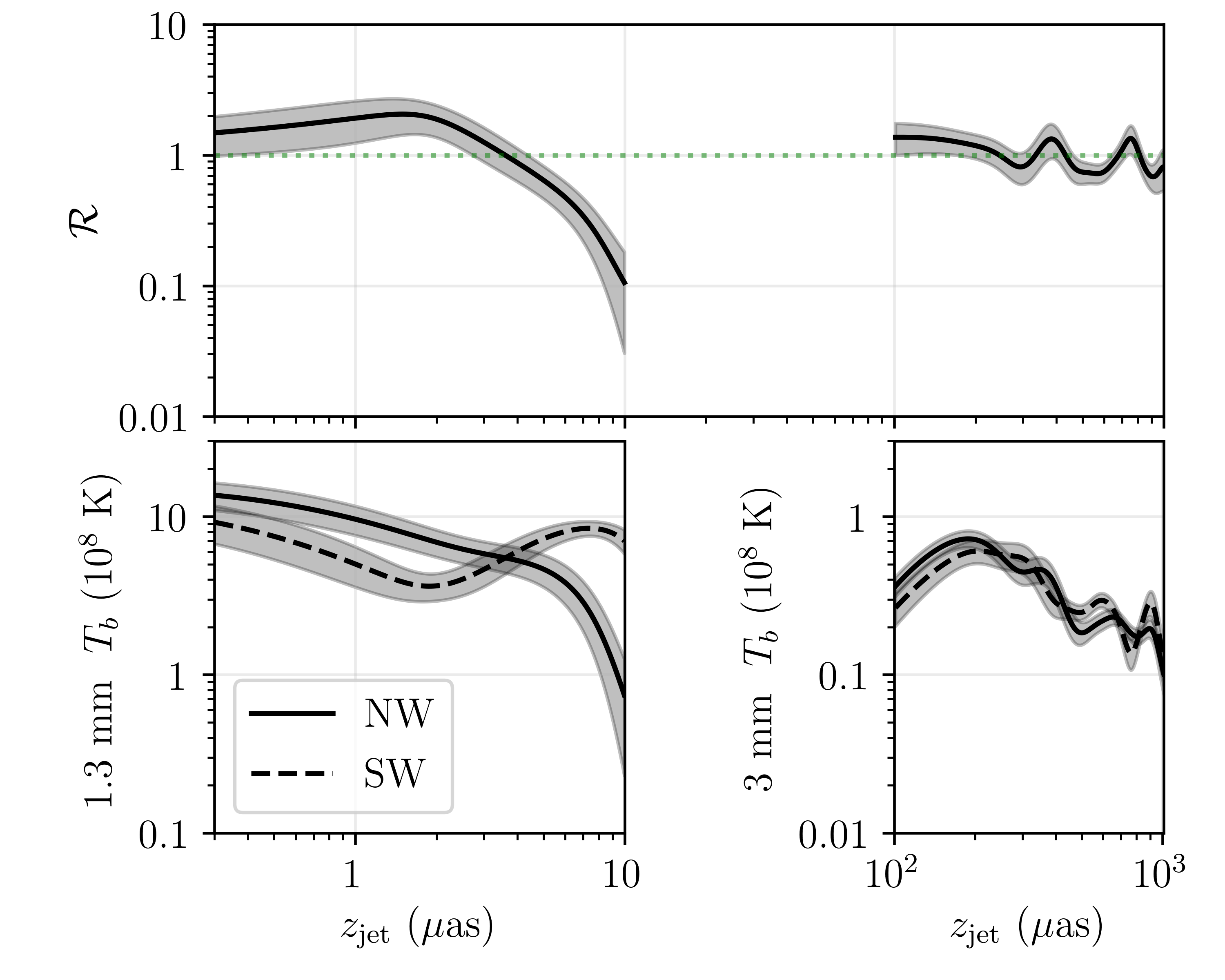}
\end{center}
\caption{Bottom: flux profiles along the top and bottom jet ridgelines shown in \autoref{fig:EHTGMVA} for the stacked 1.3~mm and 3~mm emission maps.  The latter have been smoothed over a scale of 100~\uas to remove the clear beam features.  The shaded regions indicate a combination of the intrinsic 1$\sigma$ errors and that associated with nearby ridgelines.  Top: implied brightness ratio between the two ridgelines. A brightness ratio of unity is shown by the green dotted line.}\label{fig:beaming}
\end{figure}

The dominance of the southwestern extension, in comparison to the more marginal northwestern extension, is consistent with a rapidly rotating jet structure near the black hole, aligned with the black hole spin. Brightness temperature  profiles are shown in \autoref{fig:beaming} along the corresponding paths shown in \autoref{fig:EHTGMVA}. The brightness ratio between these two components reaches values as low as ${\mathcal R}\approx0.2$ at similar projected distances from the black hole.  The uncertainty of these profiles is a combination of the intrinsic uncertainty in the reconstructions and of the ridgelines themselves; we show the uncertainty associated with a Gaussian error in the position of the parabola apex (the rightmost open green point in \autoref{fig:EHTGMVA}) of $2~\muas$ for the EHT profiles and $10~\muas$ for the 3~mm profiles.

In \autoref{fig:beaming} the difference in the brightness temperature profiles is naturally explained by black hole spin-driven jets like those first described in \citet{BZ77}. Relativistic rotational motions in the emitting plasma near the black hole, with velocities that could reach as high as $c/2$, are a consequences of the twisted magnetic field lines that penetrate the horizon and extract the rotational energy of the black hole.  These motions are seen explicitly in GRMHD simulations \citep[see, e.g.,][]{Wong2021}.  At larger distances, this rotation is suppressed due to angular momentum conservation, at which point the jet plasma motions become predominantly poloidal.  As a result, even only very modest increases in the jet height result in drastic reductions in the beaming-induced brightness asymmetry, naturally explaining its absence at 3~mm and longer wavelengths.

In both \autoref{fig:EHTGMVA} and \citetalias{M87_PaperV} the implied spin of the black hole is oriented into the sky, i.e., the black hole rotates clockwise, dragging the emitting plasma with it.  As such, this provides a striking, direct confirmation of black hole spin as the driver of the jet in \VirA.  In practice, the angular scale over which the limbs become symmetric depends on the detailed jet structure, and thus black hole spin \citep{Takahashi_etal:2018}; however, we leave further astrophysical interpretation to future work.

\section{Conclusions}
\label{sec:conclusions}
We have applied the hybrid imaging algorithm outlined in \citet{Themaging} to the 2017 EHT observations of \VirA.  This method is a Bayesian imaging and modeling scheme that reconstructs the brightness map from visibility data, accounting for station gains and atmospheric phase delays, and produces statistically meaningful posteriors for both components.  We considered both imaging and imaging with a narrow ring, finding that information criteria indicate a preference for the latter on three of four observation days.

We demonstrate that the EHT observations of the horizon scale emission of \VirA support the presence of a narrow ringlike feature.  Its radius is consistent across the seven days of the 2017 EHT observing campaign (2017 April 5-11).  The size and structure of this ring is consistent with the prominent lensed structure anticipated in horizon-resolving images of \VirA.  We associate the ring emission with the strong gravitational lensing that produces the $n=1$ photon ring; this decomposition thus represents the first direct detection of the ``back of the emission region.''  It also provides an important confirmation of the key role played by strong lensing in the formation of the images of \VirA presented in \citetalias{M87_PaperI}.

Evolving, extended diffuse emission is clearly present in addition to the bright, narrow ring.  The extended image is consistent among neighboring days, as anticipated by the long dynamical timescales in \VirA.  However, it differs from the beginning of the week to the end of the week.  This may be due either to the appearance of a new southern component or the shearing of a western component southeastward.  The latter is consistent with the direction of motion expected from the black hole spin orientation presented in \citetalias{M87_PaperV}.

The diffuse emission is dominated by compact components that surround the bright ring, similar to the morphology in many GRMHD simulation images.  This is more compact than many such images, suggesting that additional constraints on the library in \citetalias{M87_PaperV} can be made based on the compactness of this portion of the diffuse emission.

Extended components within the diffuse emission are also detected at statistically significant levels.  These include a southwestern extension, detected at between $4\sigma$ and $8\sigma$ across all days, and a northwestern extension that is only marginally detected on two days (April 5 and 11).  Both of these components are reminiscent of the larger-scale limb-brightened features seen at 3~mm; their orientation matches those extrapolated from longer wavelengths.  These may be tentatively identified with the rapidly rotating jet footprint.  The difference in the luminosity of the southern and northern components would then be naturally explained by relativistic jet rotation at the jet base.  Both of these would support the conclusion that the jet in \VirA\ is driven by the black hole rotation, as described in \citet{BZ77}.

The size of the bright ring and its relation to the diffuse emission presents a number of ways to estimate the black hole mass-to-distance ratio with varying degrees of astrophysical inputs.  All of these estimates are consistent, with a $\theta_M=4.15\pm0.74~\muas$, independent of the extent of the emission region.  Our most precise estimate arises from simultaneously reconstructing the ring and diffuse emission scales, and obtains $\theta_M=4.20^{+0.12}_{-0.06}~\muas$, which improves on the fractional uncertainty presented in \citetalias{M87_PaperVI} by a factor of more than four.  The resulting mass estimate after folding in a distance of $16.8~{\rm Mpc}$ is $7.13^{+0.20}_{-0.11}\times10^9~M_\odot$.  The uncertainty in the mass estimate is now dominated by the systematic uncertainty in the distance of $0.35\times10^9~M_\odot$.

These mass-to-distance ratio estimates are consistent with those from the variety of methods presented in \citetalias{M87_PaperVI} and from stellar dynamics \citep{Gebhardt2011}.  Because the latter is estimated from the dynamics of what are effectively test particles at distances four orders of magnitude larger than the photon orbit, the comparison of these mass measurements provide a direct test of general relativity, as described in \citetalias{M87_PaperVI}.  In practice, this test remains limited by the uncertainty in the stellar dynamics measurement.

Nevertheless, the mass estimates presented here lie at the high end of the ranges presented in \citetalias{M87_PaperVI}.  This suggests that the set of GRMHD simulations used to calibrate the mass estimates in \citetalias{M87_PaperVI} were more extended than the observed emission, biasing the calibration factor $\alpha$ toward high values and thus the mass toward low values.  This is only partly ameliorated by selecting only MAD models in the calibration process.  In comparison, no such calibration is required for two of the three mass estimates made here; in the one instance where calibration from simulations is performed, the measured systematic modification is small due to the much more robust size of the bright rings in simulated images.

Additional epochs of horizon-resolving observations will prove particularly useful in confirming the existence and nature of the bright ring.  While small variations in its location are anticipated, associated with (potentially large) variations in the location of the emission region, if the ring structure detected here is, in fact, identified with the $n=1$ photon ring, it should persist.  

The evolution of the diffuse emission map will also be diagnostic of the location and origin of the emission in \VirA.  Similar quality observations that extend over observation epochs as short as two weeks will permit the conclusive differentiation between orbiting and outflowing features \citep{Jeter2020}.  The ability to resolve dim, variable structures thus motivates longer-duration observing campaigns.

The constraints on black hole spin obtained here are inconclusive.  However, the ability to constrain $(\theta_M,a)$ to a band in the mass-spin parameter space indicates that even a single, fortuitous future EHT observation of \VirA may provide a measurement of black hole spin from gravitational lensing alone \citep{spin}.  The strength of potential spin constraints depends on the degree to which the emission region location differs during future observations.  Multiple additional measurements provides a lensing-only test of general relativity \citep{spin}.  These provide a strong motivation for including \VirA in future EHT campaigns and those of subsequent instruments.

It also suggests that future space-based millimeter-very long baseline interferometry experiments may be able to detect the next-order lensed image, i.e., that associated with the $n=2$ ring, via a similar method to that presented here \citep{spin}.  The astrophysics-independent mass measurements become substantially better constrained in this instance, and immediate tests of general relativity become possible.

Finally, this provides a direct demonstration of the ability to leverage high $S/N$ data to estimate image features with precisions that significantly exceed those implied by the ostensible observing beam.  This effective super-resolution may suggest that a similar effort applied to more distant sources may yield practical mass estimates even in the absence of resolved ring structures in images.

\mbox{}\\
\indent This work was made possible by the facilities of the Shared Hierarchical Academic Research Computing Network (SHARCNET:www.sharcnet.ca) and Compute/Calcul Canada (www.computecanada.ca).
Computations were made on the supercomputer Mammouth Parall\`ele 2 from the University of Sherbrooke, managed by Calcul Qu\'ebec and Compute Canada. The operation of this supercomputer is funded by the Canada Foundation for Innovation (CFI), the minist\`ere de l'\'Economie, de la science et de l'innovation du Qu\'ebec (MESI) and the Fonds de recherche du Qu\'ebec - Nature et technologies (FRQ-NT).
This work was supported in part by the Perimeter Institute for Theoretical Physics.  Research at the Perimeter Institute is supported by the Government of Canada through the Department of Innovation, Science and Economic Development Canada and by the Province of Ontario through the Ministry of Economic Development, Job Creation and Trade.
A.E.B. thanks the Delaney Family for their generous financial support via the Delaney Family John A. Wheeler Chair at Perimeter Institute.
A.E.B. and P.T. receive additional financial support from the Natural Sciences and Engineering Research Council of Canada through a Discovery grant. R.G.\ receives additional support from the ERC synergy grant “BlackHoleCam: Imaging the Event Horizon of Black Holes” (grant No.\ 610058). 
D.W.P. is supported by the NSF through grant Nos.\ AST-1952099, AST-1935980, AST-1828513, and AST-1440254; by the Gordon and Betty Moore Foundation through grant No.\ GBMF-5278; and in part by the Black Hole Initiative at Harvard University, which is funded by grants from the John Templeton Foundation and the Gordon and Betty Moore Foundation to Harvard University.
Furthermore, we thank Ivar Coulson for useful contributions during this project. 
H.-Y.P.\ acknowledges the support of the Ministry of Education (MoE) Yushan Young Scholar Program, the Ministry of Science and Technology (MOST) under the grant No. 110-2112-M-003-007-MY2, and National Taiwan Normal University.
I.M.V.\ acknowledges support from Research Project PID2019-108995GB-C22 of Ministerio de Ciencia e Innovaci\'on (Spain) and from the GenT Project CIDEGENT/2018/021 of Generalitat Valenciana (Spain).

\appendix

\section{Model components and priors}
\label{app:model}
We employ three model components in the comparisons made here.  These are the ``nonparameteric'' raster splined raster model, the large-scale Gaussian, and a slashed crescent.  Here we collect details of the implementation and assumed priors on each component.

\begin{deluxetable}{lccc}
\tablecaption{Hybrid Image-ring Fit Priors \label{tab:priors}}
\tablehead{
\colhead{Comp.} &
\colhead{Param.} & 
\colhead{Units} &
\colhead{Prior\tablenotemark{a}}
}
\startdata
Raster & $I_{M,N}$ & Jy~$\muas^{-2}$ & $\mathcal{L}(10^{-14},3\times10^3)$\\
I$_{M\times N}$ & FOV$_x$ & $\muas$ & $\mathcal{U}(0,200)$\\
& FOV$_y$ & $\muas$ & $\mathcal{U}(0,200)$\\
& $\phi$  & rad & $\mathcal{U}(-0.25\pi,0.25\pi)$\\
& $x$ & $\muas$ & $\mathcal{U}(-40,40)$\\
& $y$ & $\muas$ & $\mathcal{U}(-40,40)$\\
\hline
Gaussian & $I_0$ & Jy & $\mathcal{U}(0,10)$\\
A & $\sigma$ & mas & $\mathcal{U}(0.1,10^4)$\\
& $A$ & -- & $\mathcal{U}(0,1)$\\
& $\phi$ & rad & $\mathcal{U}(0,\pi)$\\
& $x$ & mas & $\mathcal{U}(-2,2)$\\
& $y$ & mas & $\mathcal{U}(-2,2)$\\
\hline
Ring & $I_0$ & Jy & $\mathcal{U}(0,2)$\\
X & $R$ & $\muas$ & $\mathcal{U}(0,10^2)$\\
& $\psi$ & -- & $\mathcal{U}(0,0.05)$\\
& $f$ & -- & $\mathcal{U}(0,1)$\\
& $\phi$ & rad & $\mathcal{U}(-\pi,\pi)$\\
& $x$ & $\muas$ & $\mathcal{U}(-40,40)$\\
& $y$ & $\muas$ & $\mathcal{U}(-40,40)$\\
\enddata
\tablenotetext{a}{Linear priors from $a$ to $b$ are represented by $\mathcal{U}(a,b)$, logarithmic priors from $a$ to $b$ are represented by $\mathcal{L}(a,b)$.}
\end{deluxetable}

\begin{figure}
\begin{center}
\includegraphics[width=\columnwidth]{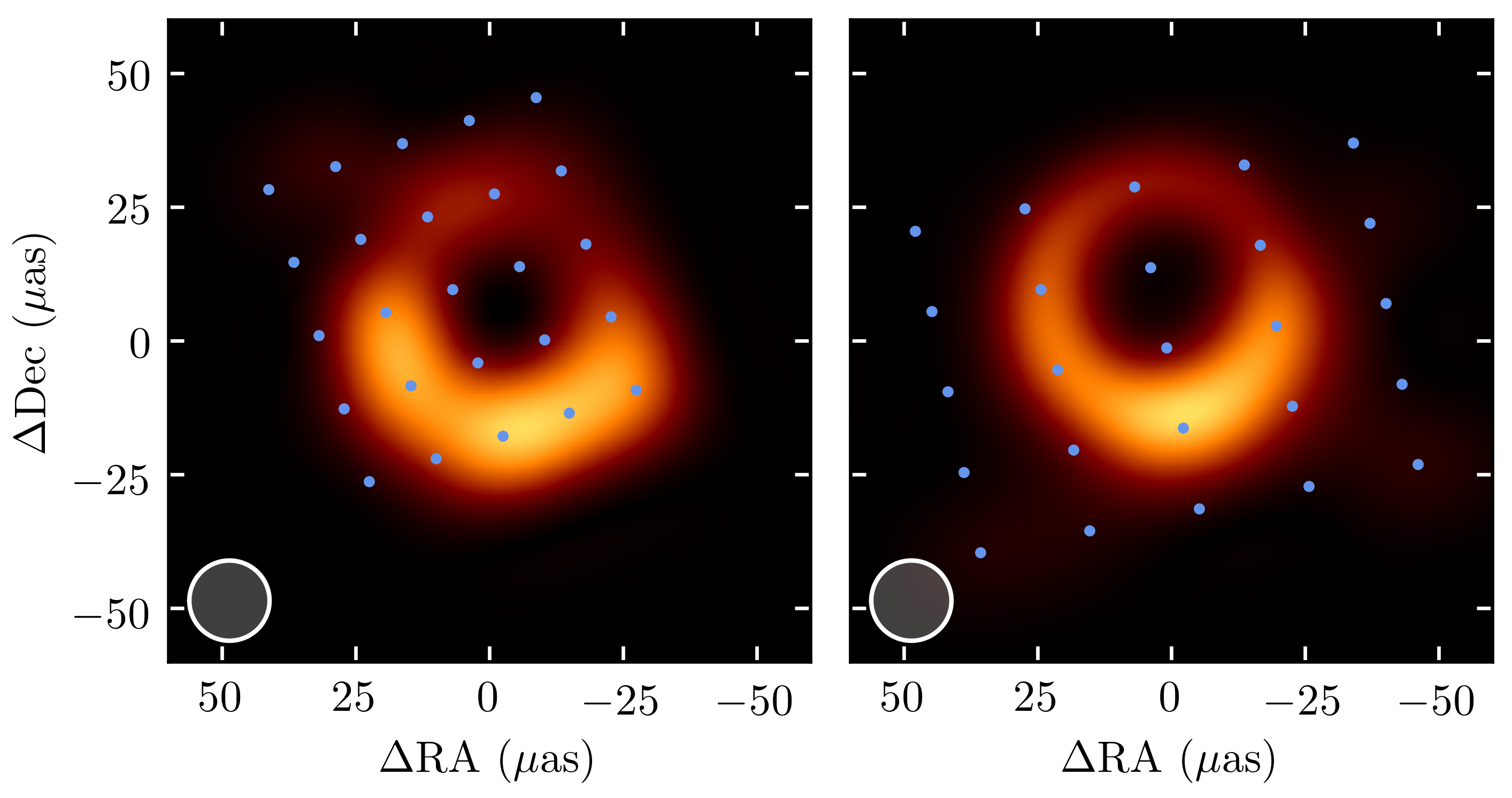}
\end{center}
\caption{Illustration of the location of image raster control points for a representative I$_{5\times5}$+A model (left) and I$_{5\times5}$+A+X model (right).  To facilitate direct comparisons between the intensity maps for the two models, both have been smoothed by a $15~\muas$ Gaussian beam, shown in the lower left of each panel.}\label{fig:raster}
\end{figure}

The splined raster model is described in \citet{Themaging}, to which we direct the reader for more detailed information.  It is characterized by a rectilinear set of control points, $I_{M,N}$, which may vary independently.  An example with the control points explicitly marked is shown in \autoref{fig:raster}.  The intensity map is interpolated between these control points using an approximate cubic spline.  We modify this by permitting a variable field of view, (FOV$_x$,FOV$_y$), orientation, $\phi$, and shift in the center of the raster, ($x$,$y$).  Parameters and priors are listed in \autoref{tab:priors}.  Note that the ring radius in angular units is given by $\theta=R(1-\psi/2)$.

The large-scale Gaussian is the asymmetric Gaussian model described in \citet{THEMIS-CODE-PAPER}.  This is characterized by a total flux, $I_0$, a symmetrized standard deviation, $\sigma$, asymmetry parameter, $A$, position angle, $\phi$, and location ($x$,$y$).  Parameters and priors are also listed in \autoref{tab:priors}.

Finally, we apply the slashed ring using the xs-ringauss model described in \citet{THEMIS-CODE-PAPER}.  We force the asymmetry parameter and additional Gaussian to vanish via the choice of appropriate priors.  This leaves the total flux, $I_0$, outer radius, $R$, fractional width, $\psi$, linear brightness gradient, $f$, brightness gradient position angle, PA, and location, ($x$,$y$).  Of these, we restrict in particular the fractional width to ensure that the ring thickness remains well below the scale of the spacings within the splined raster.

These are supplemented with a model for the scan-specific complex station gains.  The logarithmic amplitudes of the gains are assumed to have a Gaussian prior with width 20\% for all stations except the LMT, for which we assume a prior with width 100\%.  The phases priors are flat and uninformative.  The gain phases are trivially degenerate with an overall shift in the image location, and thus we remove two parameters from the total parameter count in \autoref{tab:chi2}.

\begin{figure*}
\begin{center}
\includegraphics[width=0.85\textwidth]{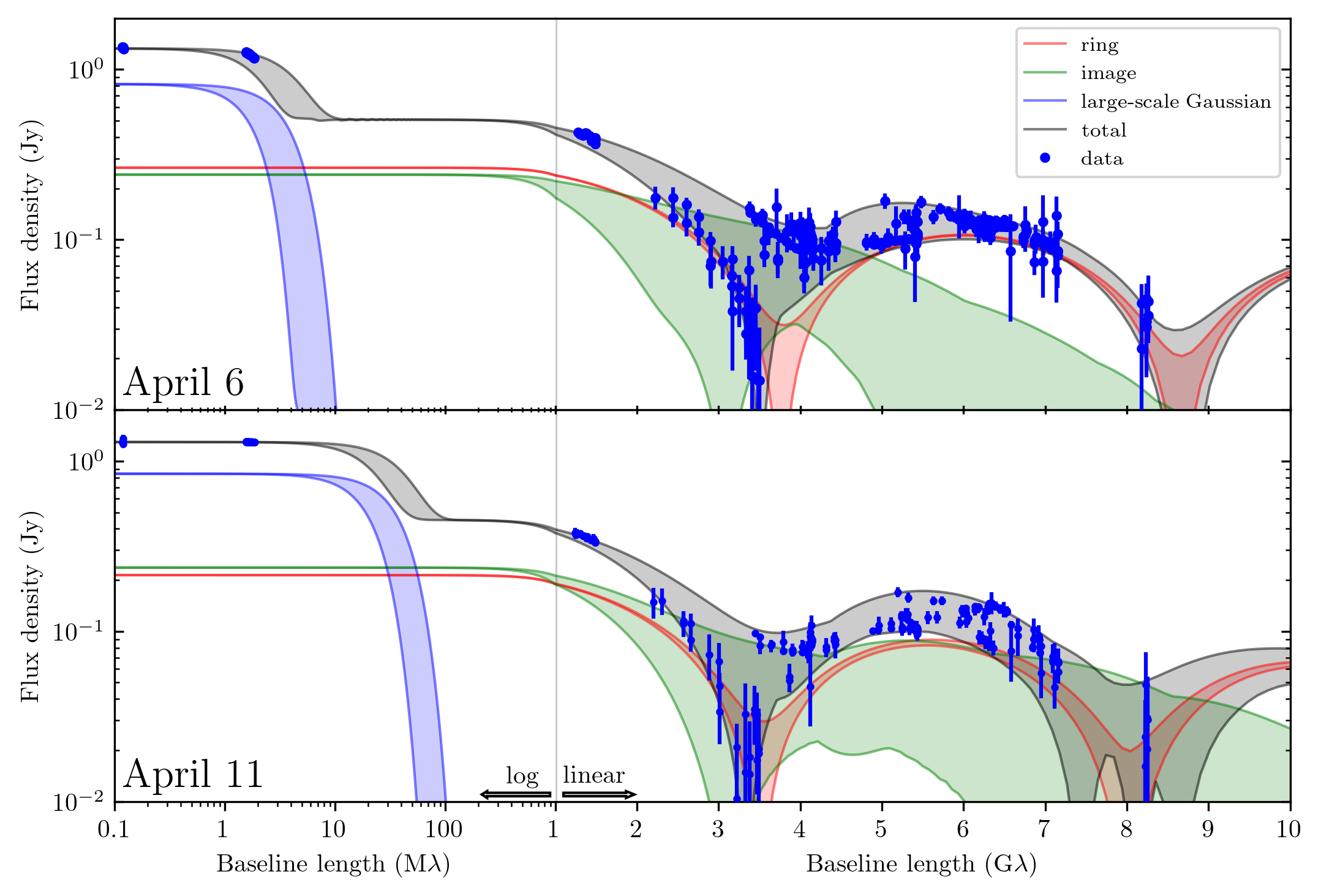}
\end{center}
\caption{Visibility amplitude versus baseline length for the April 6 (upper panel) and April 11 (lower panel) hybrid image model fits described in \autoref{sec:rings}. In both panels, the colored bands show the range of amplitudes permitted for the ring component (red), the image component (green), the large-scale Gaussian component (blue), and their sum (gray).  The gain-corrected data are plotted as blue points and error bars.}\label{fig:scale_separation}
\end{figure*}

\begin{figure}
    \centering
    \includegraphics[width=\columnwidth]{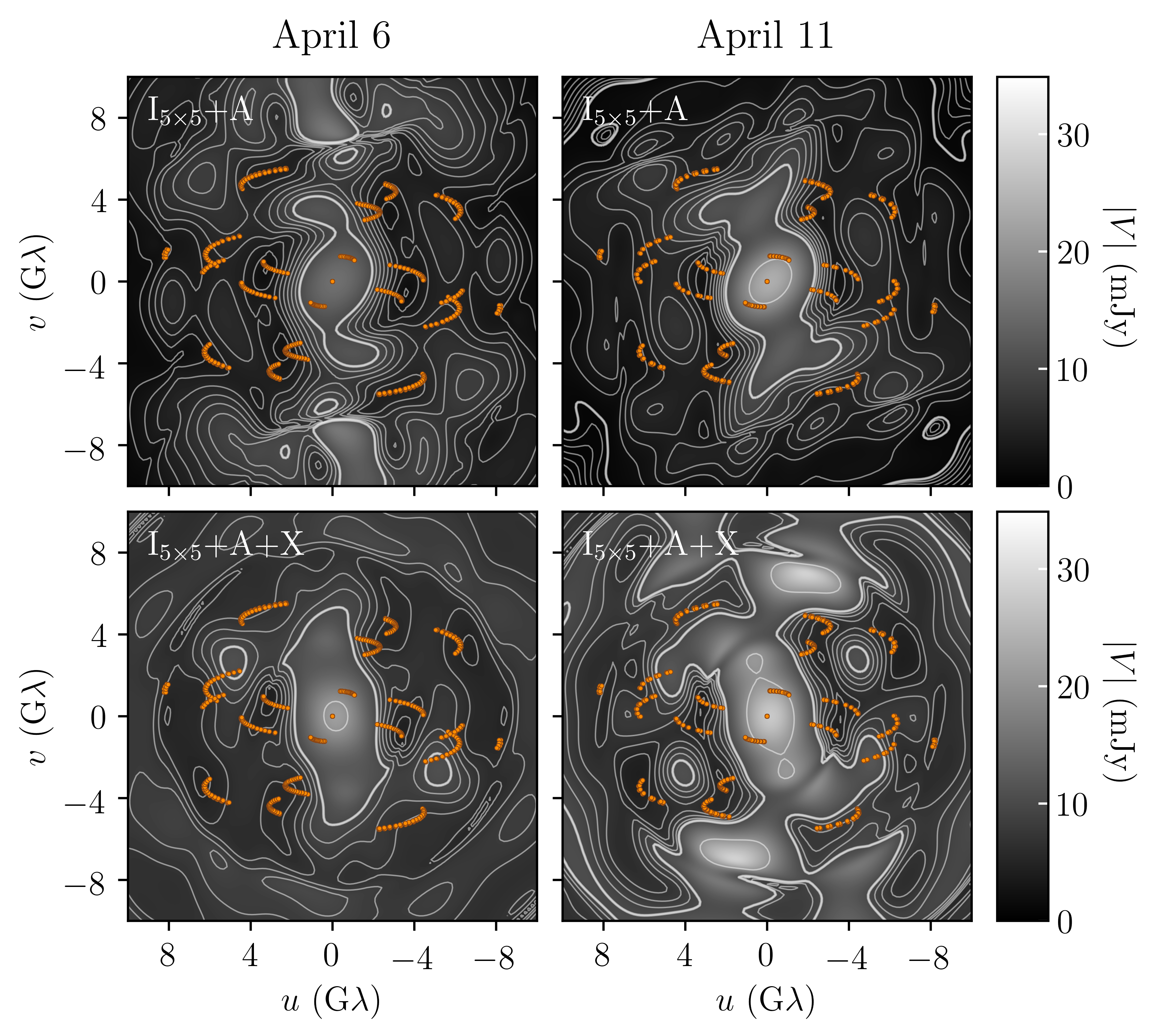}
    \caption{Posterior standard deviation maps of the visibility amplitudes on April 6 (left) and 11 (right) of the imaging reconstructions without (top) and with (bottom) a narrow ring component.  Contours are logarithimic, with thick contours on powers of 10 and thin contours at intervening $[2,3,4,5,6,7,8,9]$. \uv positions of the EHT data are shown in orange.  The color range is linear with the magnitudes indicated by the color bars on the right.}
    \label{fig:uvvar}
\end{figure}

The contribution of the individual model components are illustrated in \autoref{fig:scale_separation}.  The observed decomposition matches that anticipated by the GRMHD models explored in \citet{Themaging}.  The large-scale Gaussian only impacts the intrasite baselines, the Atacama Large Millimeter/submillimeter Array (ALMA)-Atacama Pathfinder Experiment (APEX) telescope (APEX) and the James Clerk Maxwell Telescope (JCMT)-Submillimeter Array (SMA).  The raster image and narrow ring both contribute at longer baselines, with the latter dominating at the longest baselines.  The null in the visibility amplitudes near $3.5~{\rm G}\lambda$ is reconstructed by a combination of these components, both of which exhibit nulls shifted from the observed baseline length.  In practice, \autoref{fig:scale_separation} shows the range of the visibility amplitudes at each baseline length, and thus the distinguishing power of the data exceeds that illustrated in \autoref{fig:scale_separation}.

The result of the image-reconstruction process is a posterior distribution of brightness maps, corresponding to the posterior distribution of the underlying model parameters.  As shown in \autoref{fig:uvvar} for representative days, the posterior range of visibility amplitudes associated with the image posterior is small at \uv positions near the EHT measurements and largest in the \uv holes, prominently in the north/south near $4\,{\rm G}\lambda$.  Thus, as expected, the posterior encompasses the uncertainty associated with the unknown visibilities away from the locations of the EHT data, subject to the constraints of the underlying image model.

\section{Ancillary fit results} \label{app:ancillary_posteriors}
\begin{figure*}
\includegraphics[width=\textwidth]{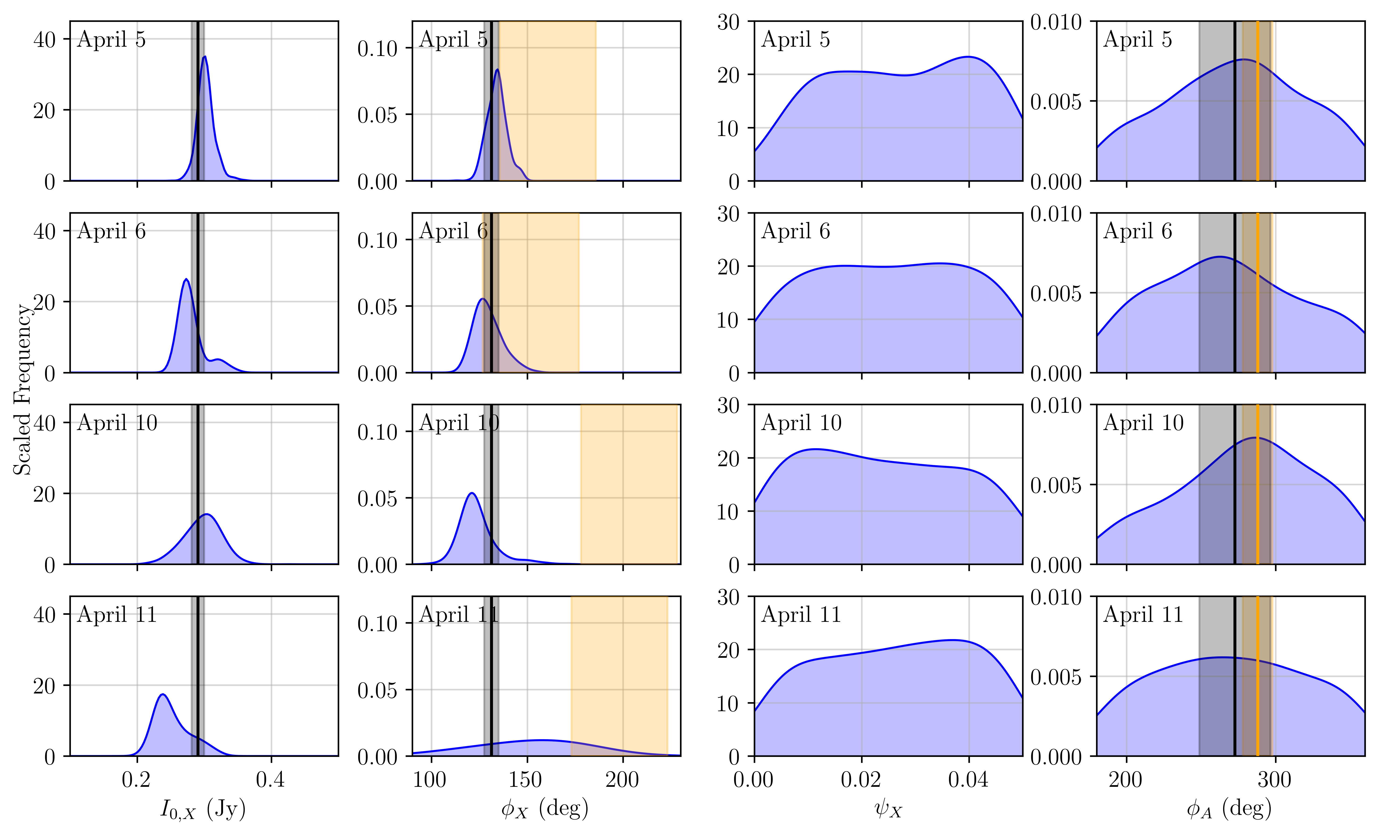}
\caption{Posterior distributions for the ring flux, position angle, fractional width ($\psi$), and position angle of the large-scale Gaussian major axis.  For the ring flux, ring positional angle, and Gaussian position angle, the black band indicates the combined $1\sigma$ estimate.  The orange bands in the ring position angle posteriors show the range of permitted values from the \themis HI+LO xs-ringauss fits reported in Table 3 of \citetalias{M87_PaperIII}.  The orange band in the Gaussian position angle posteriors indicate the orientation of the large-scale radio jet in \VirA, with a $\pm10^\circ$ range, consistent with the degree of jet precession observed.   The width posterior is indistinguishable from the prior (see \autoref{tab:priors}).} \label{fig:ancillary_posteriors}
\end{figure*}

In addition to the diameter of the bright ringlike feature, a number of additional model parameters are constrained.  In \autoref{fig:ancillary_posteriors} we collect some of the parameter constraints that may be of interest.  

The ring flux posteriors present the same information reported in \autoref{tab:ringparams} and associated discussion, to which we direct the reader for more information.  

The ring position angle estimates are consistent among days, in contrast to the reconstructions in \citetalias{M87_PaperIV} and \citetalias{M87_PaperVI}, which exhibited an evolution from April 5 and 6 to 10 and 11.  The latter change is indicated in the second column of \autoref{fig:ancillary_posteriors} by the orange shaded region.  In our analysis, this variation is entirely confined to evolution within the reconstructed diffuse emission.

The ring fractional width, $\psi$, fills the posterior, indicating that we have no traction on the width itself.

The orientation of the large-scale Gaussian is weakly constrained by the two sets of intrasite baselines, ALMA-APEX and JCMT-SMA.  The posterior of the major-axis position angle is shown in the rightmost column of \autoref{fig:ancillary_posteriors}).  These are only very weakly nonuniform, though do peak in the octant consistent with the orientation of the large-scale radio jet.  We do not attempt to interpret the large-scale Gaussian component further.

\section{\lowercase{$n=1$} ring bias estimation} \label{app:bias}
Generally, the $n=1$ photon ring is biased relative to the asymptotic $n=\infty$ photon ring, associated with the boundary of the ``shadow.''  To estimate the magnitude of this bias and its sensitivity to astrophysical uncertainties we explore both general geometric arguments and GRMHD modeling of \VirA as described in \citetalias{M87_PaperV}.

\subsection{Geometric modeling} \label{app:geometric_bias}

\begin{figure*}
\begin{center}
\includegraphics[width=0.32\textwidth]{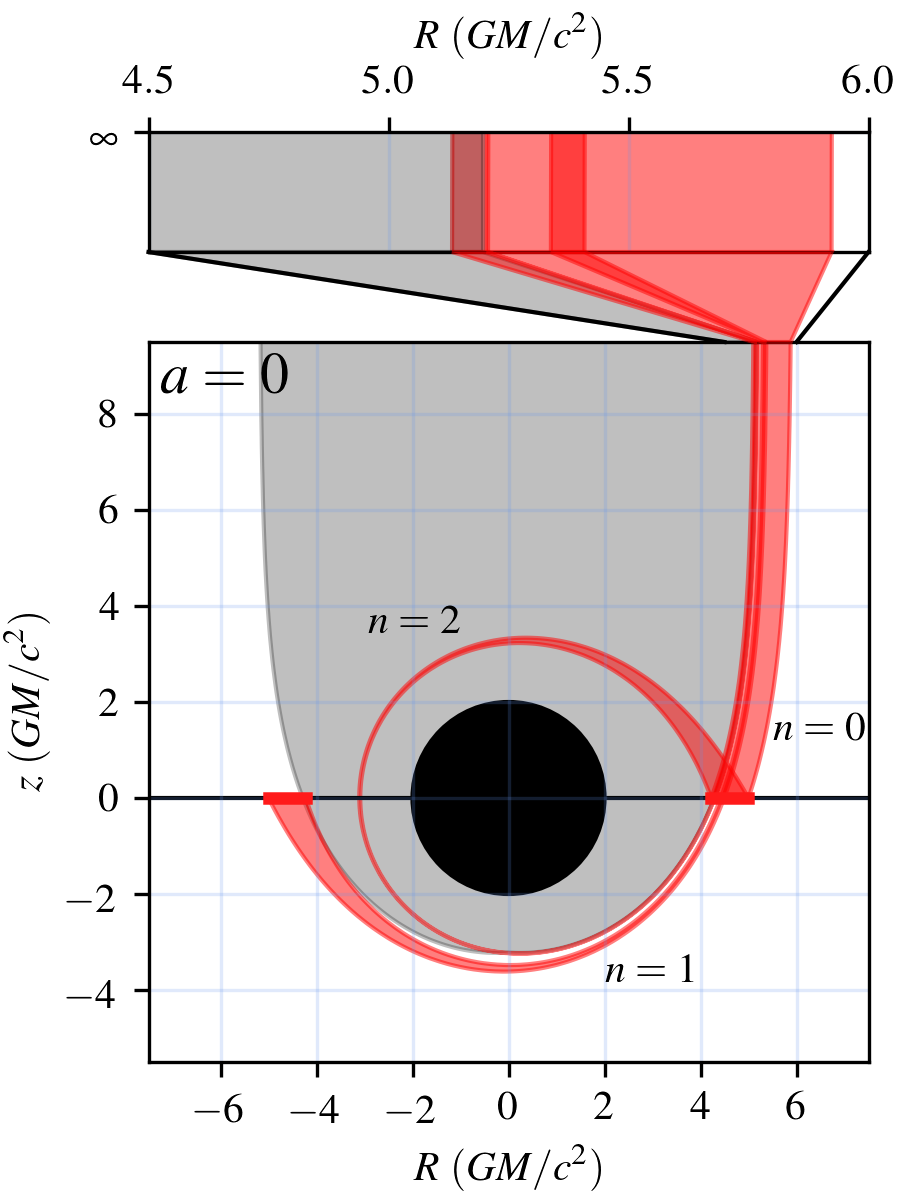}
\includegraphics[width=0.32\textwidth]{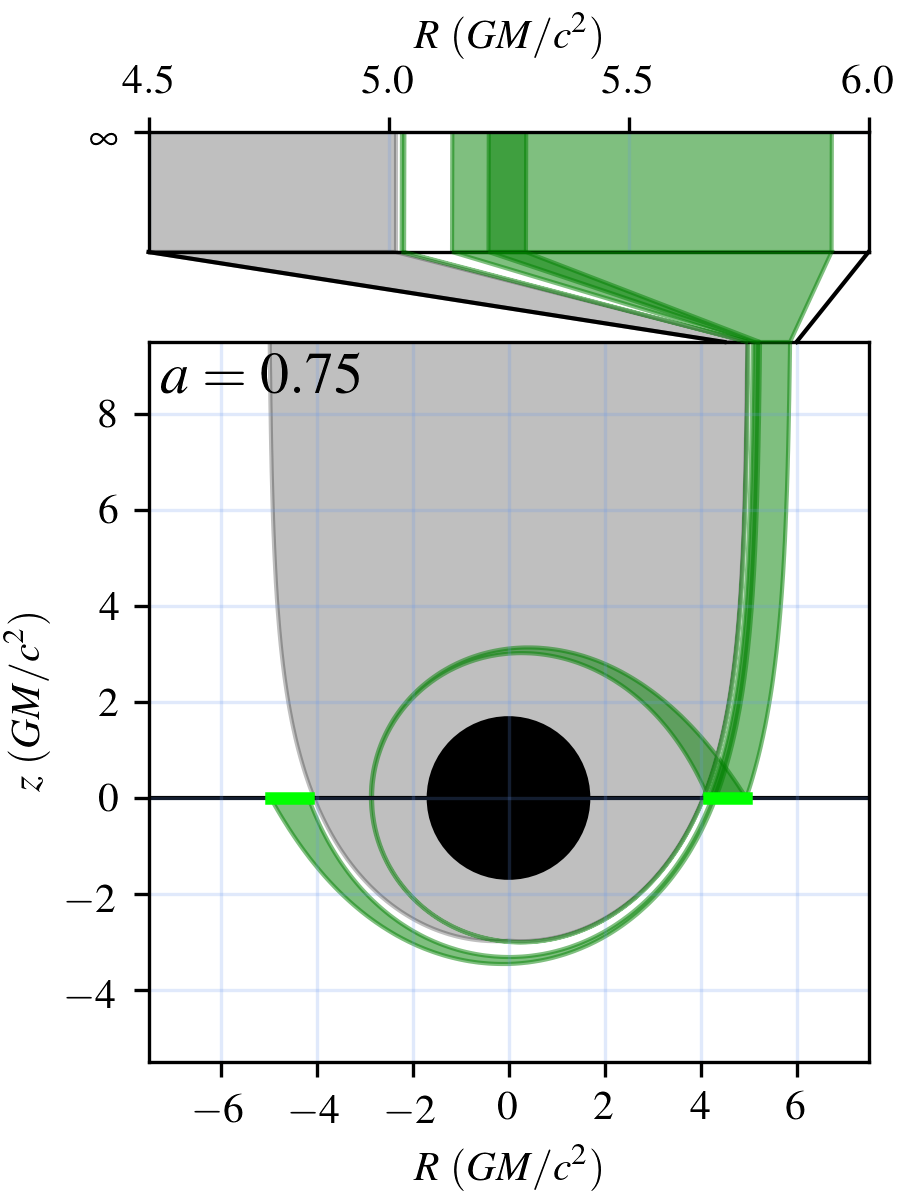}
\includegraphics[width=0.32\textwidth]{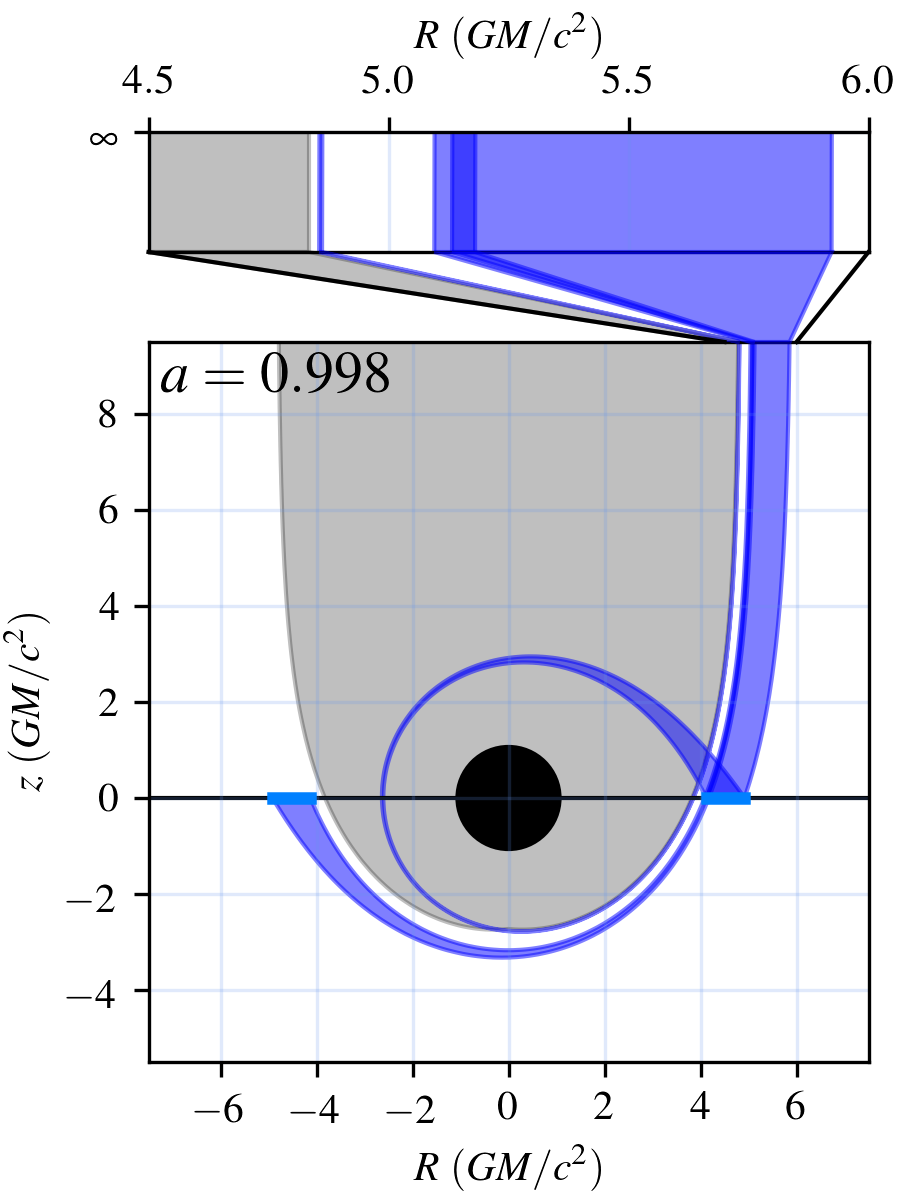}
\end{center}
\caption{Photon orbits toward a polar observer, projected into the $R\equiv r\sin\theta{\rm sgn}\phi$ and $r\cos\theta$ plane, where $r$, $\theta$, and $\phi$ are the normal Boyer-Lindquist coordinates and ${\rm sgn}\phi=\pm1$ for $|\phi|<\pi/2$ and $|\phi|>\pi/2$, respectively.  The region of the equatorial plane (thin black line) whose direct image ($n=0$) is a ring with diameter $42\pm3~\muas$ is indicated by the thick bright line segment.  The range of orbits that intersect this region for the direct emission ($n=0$), after a half-orbit and thus contributing to the $n=1$ photon ring, and after a full orbit and thus contributing to the $n=2$ photon ring, are shown.  The bar at the top shows a zoom in on the relevant side at $z=\infty$, and thus in the image plane.  For reference, the gray region corresponds to the region within the ``shadow,'' and the horizon is shown in black.  Note that an empty region can exist between the $n=0$ and $n=2$ ring simply due to the emission distribution.}\label{fig:geometric_bias_examples}
\end{figure*}

As described in \citet{Themaging}, the size of the emission region places a constraint on the magnitude of the difference between the $n=1$ and $n=\infty$ photon rings.  We compute this here by explicitly tracing rays back toward the black hole launched from a polar observer and categorize trajectories by the number of equatorial plane crossings. Relevant examples are shown in \autoref{fig:geometric_bias_examples}.  

The diameter of the observed bright ring of $42\pm3$ reported in \citetalias{M87_PaperVI} places a limit on the location of an equatorial emission region.  Assuming that this emission is dominated by the direct (i.e., $n=0$) photon trajectories, this mapping may be made explicit.  The population of orbits that terminate within the observed ring diameter is shown by the $n=0$ range in each panel of \autoref{fig:geometric_bias_examples}, and the region where they cross the equatorial plane is indicated by the thick bright line segments.  

The $n=1$ photon ring is then formed by those photon trajectories that intersect this portion of the equatorial plane after executing an additional half-orbit about the black hole.  These may be traced forward to obtain the radial position on the image plane to which these trajectories contribute, $\theta_{n=1}$.  These are also shown by the $n=1$ range in each panel of \autoref{fig:geometric_bias_examples}.  A similar range of trajectories for $n=2$ is shown for completeness.

The $n=\infty$ envelope is also computed, providing an estimate of the size of the photon ring, $\theta_{n=\infty}$.  These may then be compared, yielding an estimate of the bias, $\theta_{n=1}-\theta_{n=\infty}$.  These limits are summarized for a handful of black hole spins in \autoref{fig:geometric_bias_summary}.

In practice, due to the finite resolution of the instrument, the diameter is weakly degenerate with the subbeam substructure of the emission region.  This manifests, for example, as a dependence on the width of the ring \citepalias[see Appendix G of][]{M87_PaperIV}.  Therefore, these estimates of the range of the $n=1$ photon ring bias are overestimates of the actual bias, and thus systematic uncertainty.

\subsection{GRMHD simulation}
\begin{figure*}
\begin{center}
\includegraphics[width=0.32\textwidth]{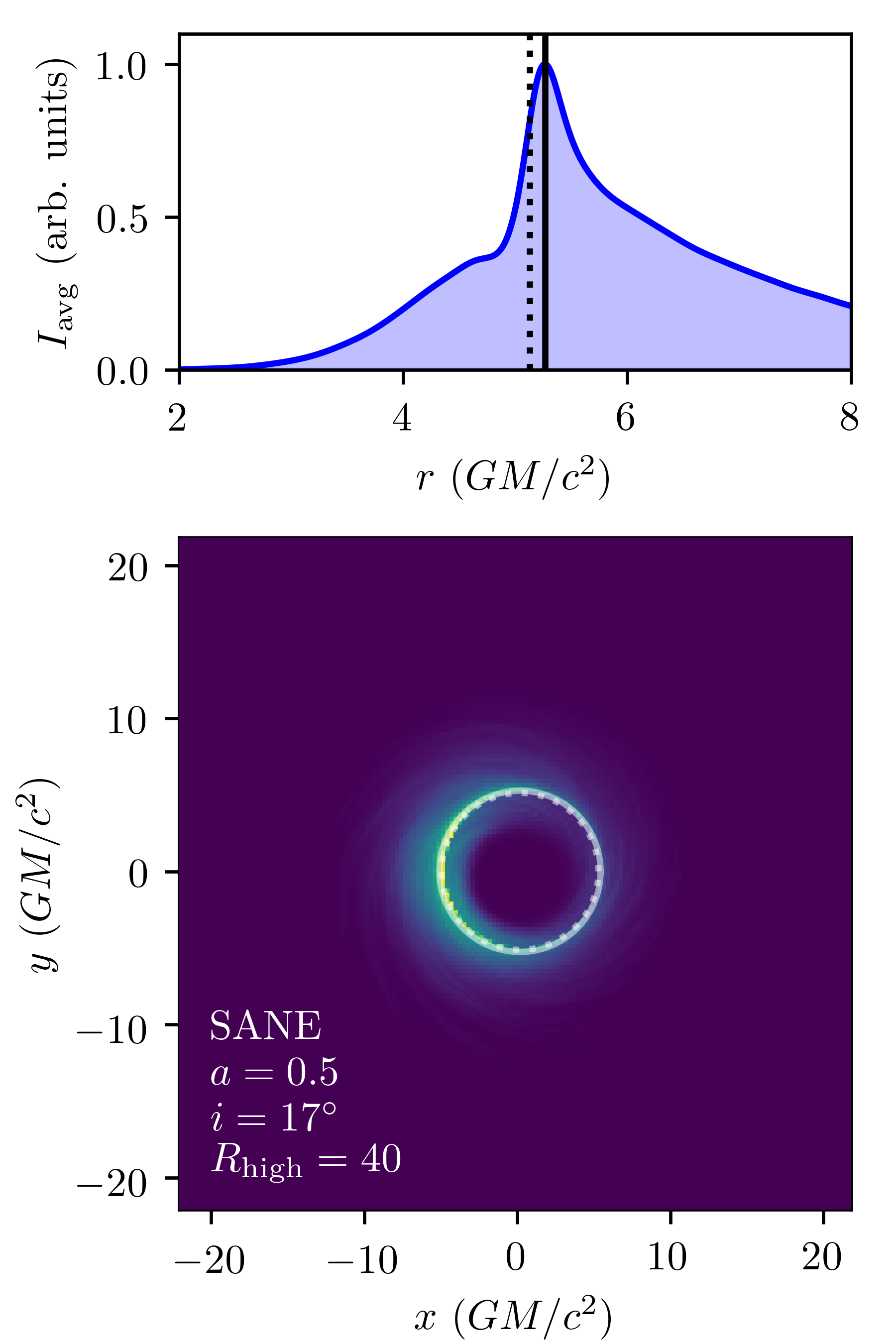}
\includegraphics[width=0.32\textwidth]{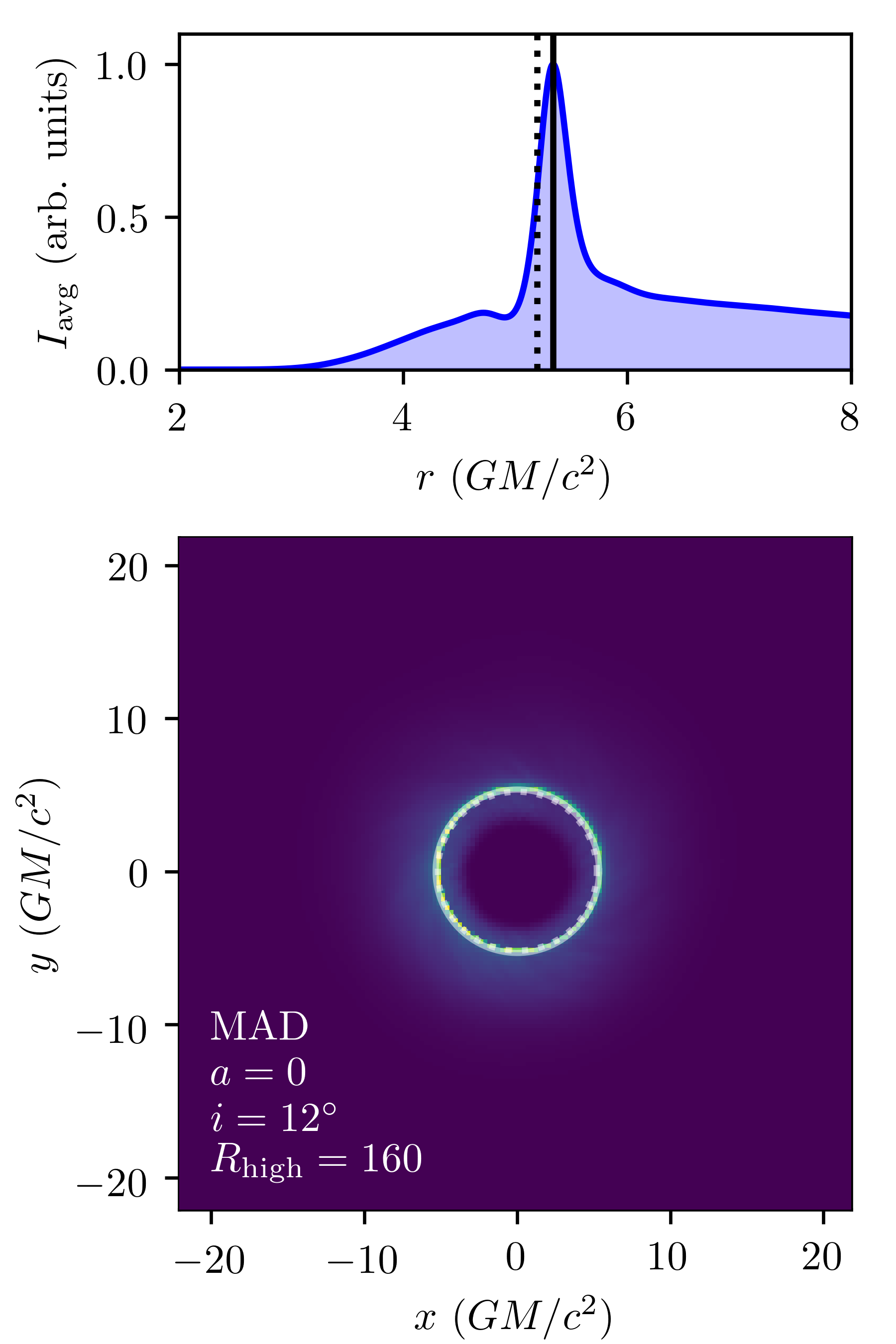}
\includegraphics[width=0.32\textwidth]{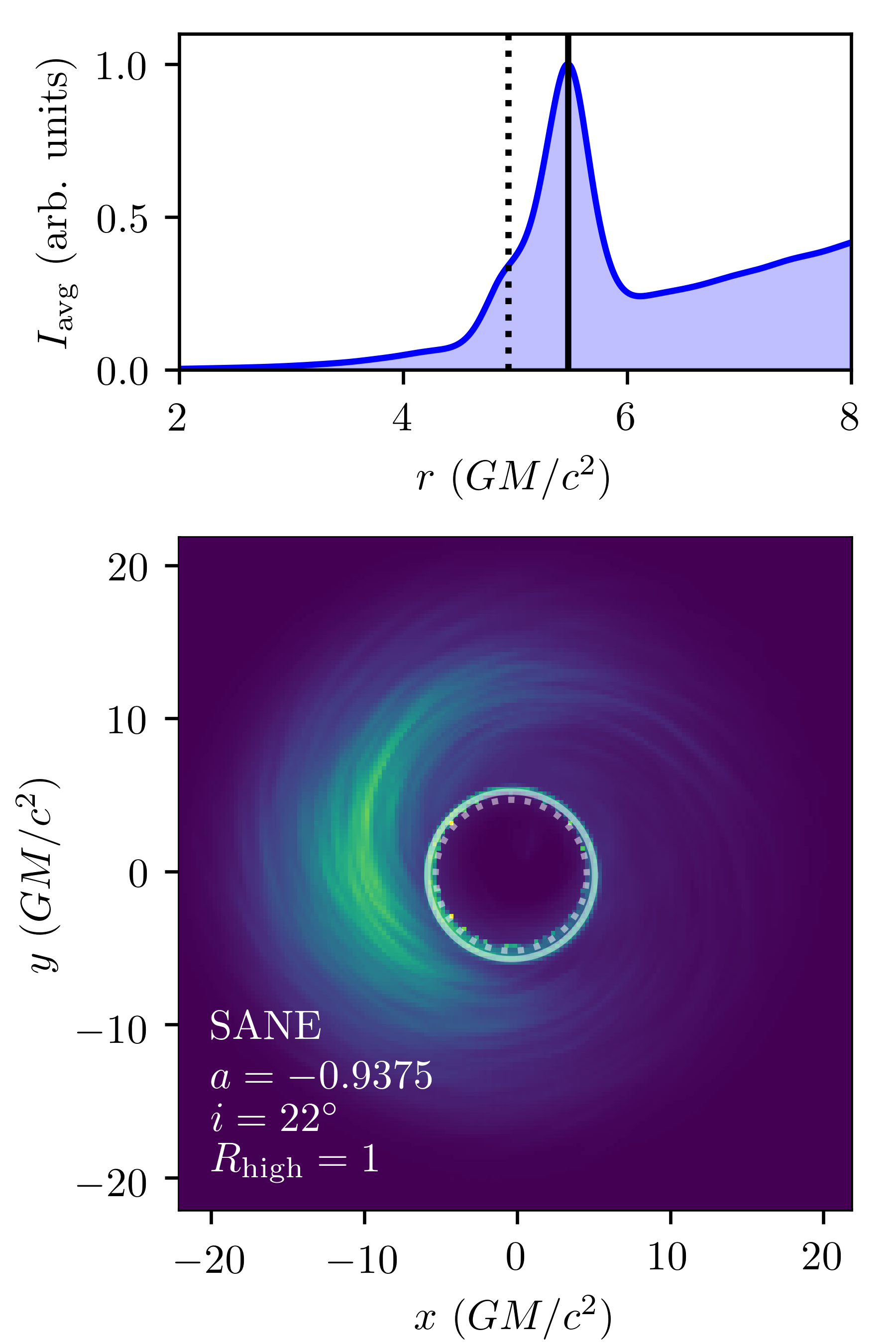}
\end{center}
\caption{Example ring radius reconstructions for three simulations, chosen at random from those reported in \citetalias{M87_PaperV}. Bottom panels show average images from each simulation.  In each, the semitransparent solid and dashed white circles indicate the best estimate of the bright ring location and the $n=\infty$ photon ring, respectively.  Top panels show the corresponding azimuthally averaged, normalized radial flux profiles.  The solid and dotted black lines indicate the best estimate of the bright ring location and the $n=\infty$ photon ring, respectively.}\label{fig:GRMHD_bias_examples}
\end{figure*}

Because the bias between the $n=1$ and $n=\infty$ photon ring radii depends on the emission distribution, we explored the relationship between the bright ring and the asymptotic photon ring in GRMHD simulation images.  We employed 198 GRMHD simulation image libraries, corresponding to those reported in \citetalias{M87_PaperV} supplemented by a set of $a=0.75$ simulations.  These span MAD and SANE morphologies,  $a\in[-0.9375,-0.5,0,0.5,0.75,0.9375]$, $i\in[12^\circ,17^\circ,22^\circ]$, and $R_{\rm high}\in[1,10,20,40,80,160]$, where negative spins correspond to counterrotating accretion flows, $R_{\rm high}$ describes the relationship between the electron and ion temperatures, and the $a=0.75$ simulations were performed only for the MAD morphology.

For each simulation, we time-averaged the simulated images to improve smoothness and reduce the impact of turbulent fluctuations.  Generally, the image center will be shifted from the image origin, i.e., $(x,y)=(0,0)$, as a result of the small nonzero inclinations and spin.  Therefore, to determine the center we minimize the variance in the radial location of the image brightness maximum between $4GM/c^2$ and $7GM/c^2$.  Finally, the brightness was computed along 36 radial chords originating from the image center, normalized by its maximum value between $2GM/c^2$ and $8GM/c^2$, and then averaged, generating $I_{\rm avg}$.  The radius of the bright ring was then taken to be the position of the peak of the $I_{\rm avg}$.  This procedure is shown for three examples in \autoref{fig:GRMHD_bias_examples}.

Each time-averaged simulation image-ring fit was visually inspected; there are no significant failures.  As anticipated by the geometric calculation in \autoref{app:geometric_bias}, the outward bias in the location of the bright ring is dependent on the extent of the emission region, growing with the second moment of the image intensity map.  The distribution of the shifts are shown in \autoref{fig:GRMHD_bias_summary} when the results for all simulations are included and when only those that were deemed acceptable in \citetalias{M87_PaperV}.

\section{Joint Mass-spin Constraint Analysis}
\label{app:massspin}
Following \citet{spin}, we model the position of the peak of the $n=0$ and $n=1$ photon rings.  The angular radii of the former are determined relative to the center of the fitted ring.  These are listed in \autoref{tab:ringparams}.

From these measurements, we construct a Gaussian log-likelihood from the measurements listed in \autoref{tab:ringparams}:
\begin{multline}
\mathcal{L}(\theta_M,a,r_{\rm em,5},r_{\rm em,6},r_{\rm em,10},r_{\rm em,11}) \\
=
\sum_{d={\rm day}}\sum_{m=0,1} \frac{
\left[\theta_{n=m,d}-\vartheta_{n=m}(a,r_{\rm em,d},i)\theta_M\right]^2}{2\sigma_{n=m,d}^2}.
\end{multline}
The six parameters are the mass, spin, and the four radii at which the emission peaks on each day.  In all cases we assume a polar observer ($i=0$); due to the weak dependence of the $n=0$ and $n=1$ photon rings on inclination near the pole and the large uncertainties on the measured ring radii, this makes a small difference.  The uncertainties $\sigma_{n=0,d}$ and $\sigma_{n=1,d}$ are computed directly from the standard deviations of the $\theta_{n=0}$ and $\theta_{n=1}$ estimates.

Priors on all parameters are adopted either on ranges large in comparison to their reconstructed values or (for spin) that encompass all physical values.  Uniform priors are adopted on the mass and spin within the intervals $[0,10~\muas]$ and $[0,1)$, respectively.  Logarithmic priors are adopted for each $r_{\rm em,d}$ on the interval $[r_h,25 r_h]$, where $r_h=1+\sqrt{1-a^2}$ is the horizon radius in units of $GM/c^2$.

We sample the corresponding posterior using the ensemble MCMC method provided by the \texttt{emcee} python package \citep{emcee2013}.  We employ 64 independent walkers and run for $10^5$ steps, discarding the first half of each chain.  Explorations with simulated data sets and inspection of the resulting MCMC chains indicate that by this time they are well converged.

\clearpage

\bibliographystyle{aasjournal_aeb}
\bibliography{references}

\begin{thebibliography}{}
\expandafter\ifx\csname natexlab\endcsname\relax\def\natexlab#1{#1}\fi
\providecommand{\url}[1]{\href{#1}{#1}}
\providecommand{\dodoi}[1]{doi:~\href{http://doi.org/#1}{\nolinkurl{#1}}}
\providecommand{\doeprint}[1]{\href{http://ascl.net/#1}{\nolinkurl{http://ascl.net/#1}}}
\providecommand{\doarXiv}[1]{\href{https://arxiv.org/abs/#1}{\nolinkurl{https://arxiv.org/abs/#1}}}

\bibitem[{{Abramowicz} \& {Fragile}(2013)}]{Abramowicz:2013}
{Abramowicz}, M.~A., \& {Fragile}, P.~C. 2013, Living Reviews in Relativity,
  16, 1

\bibitem[{{Arras} {et~al.}(2022){Arras}, {Frank}, {Haim}, {Knollm{\"u}ller},
  {Leike}, {Reinecke}, \& {En{\ss}lin}}]{Arras_2020}
{Arras}, P., {Frank}, P., {Haim}, P., {et~al.} 2022, Nature Astronomy, 6, 259

\bibitem[{{Arras} {et~al.}(2019){Arras}, {Frank}, {Leike}, {Westermann}, \&
  {En{\ss}lin}}]{Arras_2019}
{Arras}, P., {Frank}, P., {Leike}, R., {Westermann}, R., \& {En{\ss}lin}, T.~A.
  2019, \aap, 627, A134

\bibitem[{Bardeen(1973)}]{Bardeen1973}
Bardeen, J.~M. 1973, in {Black holes (Les astres occlus)}, ed. B.~S. DeWitt \&
  C.~DeWitt (New York: Gordon and Breach), 215

\bibitem[{{Blandford} \& {Znajek}(1977)}]{BZ77}
{Blandford}, R.~D., \& {Znajek}, R.~L. 1977, \mnras, 179, 433

\bibitem[{{Broderick} {et~al.}(2020{\natexlab{a}}){Broderick}, {Gold},
  {et~al.}}]{THEMIS-CODE-PAPER}
{Broderick}, A., {Gold}, R., {et~al.} 2020{\natexlab{a}}, \apj, 897, 139

\bibitem[{{Broderick} \& {Loeb}(2009)}]{Broderick2009}
{Broderick}, A.~E., \& {Loeb}, A. 2009, \apjl, 703, L104

\bibitem[{{Broderick} {et~al.}(2020{\natexlab{b}}){Broderick}, {Pesce},
  {Tiede}, {Pu}, \& {Gold}}]{Themaging}
{Broderick}, A.~E., {Pesce}, D.~W., {Tiede}, P., {Pu}, H.-Y., \& {Gold}, R.
  2020{\natexlab{b}}, \apj, 898, 9

\bibitem[{{Broderick} {et~al.}(2022){Broderick}, {Tiede}, {Pesce}, \&
  {Gold}}]{spin}
{Broderick}, A.~E., {Tiede}, P., {Pesce}, D.~W., \& {Gold}, R. 2022, \apj, 927,
  6

\bibitem[{Broderick {et~al.}(2016)Broderick, Fish, Johnson, Rosenfeld, Wang,
  Doeleman, Akiyama, Johannsen, \& Roy}]{Broderick2016}
Broderick, A.~E., Fish, V.~L., Johnson, M.~D., {et~al.} 2016, \apj, 820, 137

\bibitem[{{Carilli} \& {Thyagarajan}(2022)}]{Carilli_2021}
{Carilli}, C.~L., \& {Thyagarajan}, N. 2022, \apj, 924, 125

\bibitem[{{Carpenter} {et~al.}(2017){Carpenter}, {Gelman}, {Hoffman}, {Lee},
  {Goodrich}, {Betancourt}, {Brubaker}, {Guo}, {Li}, \& {Riddell}}]{Stan:2017}
{Carpenter}, B., {Gelman}, A., {Hoffman}, M.~D., {et~al.} 2017, Journal of
  Statistical Software, 76, 1

\bibitem[{{Chael} {et~al.}(2019){Chael}, {Narayan}, \& {Johnson}}]{Chael19}
{Chael}, A., {Narayan}, R., \& {Johnson}, M.~D. 2019, \mnras, 486, 2873

\bibitem[{{Davelaar} {et~al.}(2019){Davelaar}, {Olivares}, {Porth},
  {Bronzwaer}, {Janssen}, {Roelofs}, {Mizuno}, {Fromm}, {Falcke}, \&
  {Rezzolla}}]{Davelaar19}
{Davelaar}, J., {Olivares}, H., {Porth}, O., {et~al.} 2019, \aap, 632, A2

\bibitem[{{Event Horizon Telescope Collaboration}
  {et~al.}(2019{\natexlab{a}})}]{M87_PaperI}
{Event Horizon Telescope Collaboration}, {et~al.} 2019{\natexlab{a}}, \apjl,
  875, L1

\bibitem[{{Event Horizon Telescope Collaboration}
  {et~al.}(2019{\natexlab{b}})}]{M87_PaperII}
---. 2019{\natexlab{b}}, \apjl, 875, L2

\bibitem[{{Event Horizon Telescope Collaboration}
  {et~al.}(2019{\natexlab{c}})}]{M87_PaperIII}
---. 2019{\natexlab{c}}, \apjl, 875, L3

\bibitem[{{Event Horizon Telescope Collaboration}
  {et~al.}(2019{\natexlab{d}})}]{M87_PaperIV}
---. 2019{\natexlab{d}}, \apjl, 875, L4

\bibitem[{{Event Horizon Telescope Collaboration}
  {et~al.}(2019{\natexlab{e}})}]{M87_PaperV}
---. 2019{\natexlab{e}}, \apjl, 875, L5

\bibitem[{{Event Horizon Telescope Collaboration}
  {et~al.}(2019{\natexlab{f}})}]{M87_PaperVI}
---. 2019{\natexlab{f}}, \apjl, 875, L6

\bibitem[{{Foreman-Mackey} {et~al.}(2013){Foreman-Mackey}, {Hogg}, {Lang}, \&
  {Goodman}}]{emcee2013}
{Foreman-Mackey}, D., {Hogg}, D.~W., {Lang}, D., \& {Goodman}, J. 2013, \pasp,
  125, 306

\bibitem[{{Gammie} {et~al.}(2003){Gammie}, {McKinney}, \&
  {T{\'o}th}}]{Gammie2003}
{Gammie}, C.~F., {McKinney}, J.~C., \& {T{\'o}th}, G. 2003, \apj, 589, 444

\bibitem[{{Gebhardt} {et~al.}(2011){Gebhardt}, {Adams}, {Richstone}, {Lauer},
  {Faber}, {G{\"u}ltekin}, {Murphy}, \& {Tremaine}}]{Gebhardt2011}
{Gebhardt}, K., {Adams}, J., {Richstone}, D., {et~al.} 2011, \apj, 729, 119

\bibitem[{{Gralla} {et~al.}(2019){Gralla}, {Holz}, \& {Wald}}]{Gralla_2019}
{Gralla}, S.~E., {Holz}, D.~E., \& {Wald}, R.~M. 2019, \prd, 100, 024018

\bibitem[{{Hada} {et~al.}(2011){Hada}, {Doi}, {Kino}, {Nagai}, {Hagiwara}, \&
  {Kawaguchi}}]{Hada2011}
{Hada}, K., {Doi}, A., {Kino}, M., {et~al.} 2011, \nat, 477, 185

\bibitem[{{Jeter} \& {Broderick}(2021)}]{Jeter2020b}
{Jeter}, B., \& {Broderick}, A.~E. 2021, \apj, 908, 139

\bibitem[{{Jeter} {et~al.}(2020){Jeter}, {Broderick}, \& {Gold}}]{Jeter2020}
{Jeter}, B., {Broderick}, A.~E., \& {Gold}, R. 2020, \mnras, 493, 5606

\bibitem[{{Jeter} {et~al.}(2019){Jeter}, {Broderick}, \&
  {McNamara}}]{Jeter2019}
{Jeter}, B., {Broderick}, A.~E., \& {McNamara}, B.~R. 2019, \apj, 882, 82

\bibitem[{{Johnson} {et~al.}(2020){Johnson}, {Lupsasca}, {Strominger}, {Wong},
  {Hadar}, {Kapec}, {Narayan}, {Chael}, {Gammie}, {Galison}, {Palumbo},
  {Doeleman}, {Blackburn}, {Wielgus}, {Pesce}, {Farah}, \&
  {Moran}}]{Johnson_2019}
{Johnson}, M.~D., {Lupsasca}, A., {Strominger}, A., {et~al.} 2020, Science
  Advances, 6, eaaz1310

\bibitem[{{Kim} {et~al.}(2018){Kim}, {Krichbaum}, {Lu}, {Ros}, {Bach},
  {Bremer}, {de Vicente}, {Lindqvist}, \& {Zensus}}]{M873mm}
{Kim}, J.~Y., {Krichbaum}, T.~P., {Lu}, R.~S., {et~al.} 2018, \aap, 616, A188

\bibitem[{{Mo{\'s}cibrodzka} {et~al.}(2016){Mo{\'s}cibrodzka}, {Falcke}, \&
  {Shiokawa}}]{Moscibrodzka2016}
{Mo{\'s}cibrodzka}, M., {Falcke}, H., \& {Shiokawa}, H. 2016, \aap, 586, A38

\bibitem[{{Pesce}(2021)}]{DMC}
{Pesce}, D.~W. 2021, \aj, 161, 178

\bibitem[{{Porth} {et~al.}(2019){Porth}, {Chatterjee}, {Narayan}, \& {Event
  Horizon Telescope Collaboration}}]{Porth2019}
{Porth}, O., {Chatterjee}, K., {Narayan}, R., \& {Event Horizon Telescope
  Collaboration}. 2019, \apjs, 243, 26

\bibitem[{{Sun} \& {Bouman}(2020)}]{Sun_2020}
{Sun}, H., \& {Bouman}, K.~L. 2020, arXiv e-prints, arXiv:2010.14462.
\newblock \doarXiv{2010.14462}

\bibitem[{{Syed} {et~al.}(2019){Syed}, {Bouchard-C{\^o}t{\'e}},
  {Deligiannidis}, \& {Doucet}}]{DEO:2019}
{Syed}, S., {Bouchard-C{\^o}t{\'e}}, A., {Deligiannidis}, G., \& {Doucet}, A.
  2019, arXiv e-prints, arXiv:1905.02939.
\newblock \doarXiv{1905.02939}

\bibitem[{{Takahashi} {et~al.}(2018){Takahashi}, {Toma}, {Kino}, {Nakamura}, \&
  {Hada}}]{Takahashi_etal:2018}
{Takahashi}, K., {Toma}, K., {Kino}, M., {Nakamura}, M., \& {Hada}, K. 2018,
  \apj, 868, 82

\bibitem[{{Thompson} {et~al.}(2017){Thompson}, {Moran}, \& {Swenson}}]{TMS}
{Thompson}, A.~R., {Moran}, J.~M., \& {Swenson}, Jr., G.~W. 2017,
  {Interferometry and Synthesis in Radio Astronomy, 3rd Edition} (Springer)

\bibitem[{Tiede(2021)}]{TiedeThesis}
Tiede, P. 2021, PhD thesis, University of Waterloo

\bibitem[{{Walker} {et~al.}(2018){Walker}, {Hardee}, {Davies}, {Ly}, \&
  {Junor}}]{Walker_2018}
{Walker}, R.~C., {Hardee}, P.~E., {Davies}, F.~B., {Ly}, C., \& {Junor}, W.
  2018, \apj, 855, 128

\bibitem[{{Walsh} {et~al.}(2013){Walsh}, {Barth}, {Ho}, \& {Sarzi}}]{Walsh2013}
{Walsh}, J.~L., {Barth}, A.~J., {Ho}, L.~C., \& {Sarzi}, M. 2013, \apj, 770, 86

\bibitem[{{Wong} {et~al.}(2021){Wong}, {Du}, {Prather}, \& {Gammie}}]{Wong2021}
{Wong}, G.~N., {Du}, Y., {Prather}, B.~S., \& {Gammie}, C.~F. 2021, \apj, 914,
  55

\end{thebibliography}

\end{document}